\documentclass[8.5pt,twoside,twocolumn]{article}
\usepackage[utf8]{inputenc}
\usepackage[super,sort&compress,comma]{natbib} 
\usepackage{mhchem}
\usepackage{times,mathptmx}
\usepackage{sectsty}
\usepackage{balance} 
\usepackage{amsmath,amssymb}

\usepackage{graphicx} %
\usepackage{lastpage}
\usepackage[format=plain,justification=raggedright,singlelinecheck=false,font=small,labelfont=bf,labelsep=space]{caption} 
\usepackage{fancyhdr}
\usepackage{url}

\graphicspath{
	{./images/}
}

\usepackage{xcolor}
\usepackage{prettyref}
\newrefformat{fig}{Fig.\,\ref{#1}}
\newrefformat{sec}{sect.\,\ref{#1}}
\usepackage{nicefrac}
\usepackage{siunitx}

\newcommand{\re}{\mathsf{Re}}
\newcommand{\la}{\mathsf{La}}
\newcommand{\ca}{\mathsf{Ca}}
\newcommand{\ma}{\mathsf{Ma}}
\newcommand{\dint}{\mathrm{d}}

\begin{document}
\thispagestyle{plain}
\fancypagestyle{plain}{
\fancyhead[L]{\includegraphics[height=8pt]{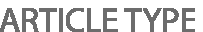}}
\fancyhead[C]{\hspace{-1cm}\includegraphics[height=20pt]{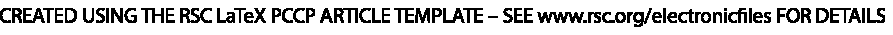}}
\fancyhead[R]{\includegraphics[height=10pt]{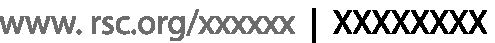}\vspace{-0.2cm}}
\renewcommand{\headrulewidth}{1pt}}
\renewcommand{\thefootnote}{\fnsymbol{footnote}}
\renewcommand\footnoterule{\vspace*{1pt}%
\hrule width 3.4in height 0.4pt \vspace*{5pt}} 
\setcounter{secnumdepth}{5}

\makeatletter 
\def\subsubsection{\@startsection{subsubsection}{3}{10pt}{-1.25ex plus -1ex minus -.1ex}{0ex plus 0ex}{\normalsize\bf}} 
\def\paragraph{\@startsection{paragraph}{4}{10pt}{-1.25ex plus -1ex minus -.1ex}{0ex plus 0ex}{\normalsize\textit}} 
\renewcommand\@biblabel[1]{#1}            
\renewcommand\@makefntext[1]%
{\noindent\makebox[0pt][r]{\@thefnmark\,}#1}
\makeatother 
\renewcommand{\figurename}{\small{Fig.}~}
\sectionfont{\large}
\subsectionfont{\normalsize} 

\fancyfoot{}
\fancyfoot[LO,RE]{\vspace{-7pt}\includegraphics[height=9pt]{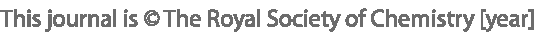}}
\fancyfoot[CO]{\vspace{-7.2pt}\hspace{12.2cm}\includegraphics{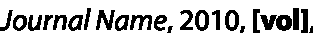}}
\fancyfoot[CE]{\vspace{-7.5pt}\hspace{-13.5cm}\includegraphics{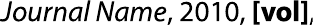}}
\fancyfoot[RO]{\footnotesize{\sffamily{1--\pageref{LastPage} ~\textbar  \hspace{2pt}\thepage}}}
\fancyfoot[LE]{\footnotesize{\sffamily{\thepage~\textbar\hspace{3.45cm} 1--\pageref{LastPage}}}}
\fancyhead{}
\renewcommand{\headrulewidth}{1pt} 
\renewcommand{\footrulewidth}{1pt}
\setlength{\arrayrulewidth}{1pt}
\setlength{\columnsep}{6.5mm}
\setlength\bibsep{1pt}

\twocolumn[
  \begin{@twocolumnfalse}
\noindent\LARGE{\textbf{Inertial migration and axial control of deformable capsules $^\dag$}}
\vspace{0.6cm}

\noindent\large{\textbf{Christian Schaaf$^{\ast}$$^{a}$} and
Holger Stark{$^{a}$}}\vspace{0.5cm}

\noindent\textit{\small{\textbf{Received Xth XXXXXXXXXX 20XX, Accepted Xth XXXXXXXXX 20XX\newline
First published on the web Xth XXXXXXXXXX 200X}}}

\noindent \textbf{\small{DOI: 10.1039/b000000x}}
\vspace{0.6cm}

\noindent \normalsize{The mechanical deformability of single cells 
is
an important indicator for various diseases 
such as
cancer, blood diseases and inflammation. 
Lab-on-a-chip devices allow
to 
separate such cells
from healthy cells
using hydrodynamic 
forces.
We perform hydrodynamic simulations based on the lattice-Boltzmann method
and
study the behavior of an elastic capsule in 
a
microfluidic channel flow
in the inertial regime.
While 
inertial lift forces drive
the 
capsule
away from the channel center, 
its
deformability
favors
migration in the opposite direction.
Balancing both
migration mechanisms,
a deformable capsule
assembles at a specific equilibrium distance depending on its size and deformability.
We find that this equilibrium distance is 
nearly
independent 
of
the channel Reynolds number
and falls on a 
single master curve 
when plotted versus the Laplace number. We identify a similar 
master curve for varying particle radius.
In contrast, the actual deformation of
a
capsule strongly depends on the Reynolds 
number. The lift-force profiles behave in a similar manner as those for rigid particles.
Using the Saffman effect, the capsule's
equilibrium position can be 
controled
by an external 
force
along the channel axis.
While rigid particles move to the center when slowed down, very soft capsules 
show the opposite behavior. Interestingly, for
a specific control force 
particles are
focused 
on the same equilibrium position
independent of 
their
deformability.
	}
\vspace{0.5cm}
\end{@twocolumnfalse}
]
  
\footnotetext{\textit{$^{a}$~Institute of Theoretical Physics, Technical University Berlin, Hardenbergstr. 36, 10623 Berlin, Germany; E-mail: Christian.Schaaf@tu-berlin.de}}

\section{Introduction}

The mechanical 
deformability
of 
biological
cells 
is a sensitive
quantity 
for identifying
various diseases\cite{lee_biomechanics_2007}. Tumor cells, for example, 
are
much softer than healthy cells\cite{suresh_biomechanics_2007,guck_optical_2005}, while 
malaria-infected red blood cells 
become
much stiffer
when occupied by the parasite
\cite{cranston_plasmodium_1984,suresh_connections_2005}. Microfluidic lab-on-a-chip devices utilize 
hydrodynamic effects 
in order to
separate and probe
biological cell samples\cite{otto_real-time_2015,hou_deformability_2010,huang_continuous_2004,geislinger_separation_2012}. 
Such lab-on-a-chip devices are 
easy to use and can be designed as mass product.
They
are thus a helpful tool 
for especially improving
medical conditions 
in poor countries\cite{bell_ensuring_2006}. 

The dynamics of deformable particles such as vesicles\cite{coupier_noninertial_2008}, capsules\cite{risso_experimental_2006,bagchi_mesoscale_2007},
or red blood cells\cite{lazaro_rheology_2014,bacher_clustering_2017}
has
mostly
been
studied
at low Reynolds numbers. Deformable particles migrate towards regions of low shear gradient. Thus, when
immersed in a Poiseuille flow 
they move
towards the channel center
\cite{kaoui_lateral_2008}. 

At medium Reynolds numbers
Segr\'e and Silberberg observed that rigid particles in 
Poiseuille
flow 
assemble 
in
an annulus 
roughly
half way between the 
center 
and the wall of a channel with circular cross section
\cite{segre_radial_1961,di_carlo_inertial_2009}. 
Thus, 
in the non-uniform
flow profile 
an inertial lift force occurs that
initiates a drift motion directed away from the channel center.
At the equilibrium distance this force is compensated by repulsive particle-wall 
interaction.
The Segr\'e-Silberberg
effect is well studied in experiments\cite{matas_lateral_2004,carlo_continuous_2007,bhagat_inertial_2009}, in theory using matched asymptotics \cite{asmolov_inertial_1999,schonberg_inertial_1989}, and in simulations\cite{chun_inertial_2006,prohm_inertial_2012}. While the original 
setup
used a cylindrical channel, 
microfluidic
experiments are usually conducted with 
quadratic or rectangular channels, as they are easier to manufacture. 
Due to the
reduced symmetry,
the circular annulus is replaced by
eight off-centered equilibrium positions in case of a quadratic channel (four on the main axes and four on the diagonals) or
two 
on the short axes
of a rectangular channel\cite{lee_dynamic_2010,prohm_feedback_2014}.
In
recent years also the dynamics of deformable particles in the inertial regime 
has been
studied \cite{salac_reynolds_2012,mach_continuous_2010,doddi_lateral_2008,sun_numerical_2015,kim_inertial_2015,wang_motion_2016}. 
In particular, \citet{hur_deformability-based_2011} demonstrated
with their experiments that particles can be separated 
from each other
based on their 
elastic deformability. Indeed,
particles move closer to the channel center the softer and 
also
the larger they 
are.
This effect was also studied in simulations\cite{kilimnik_inertial_2011, chen_inertia-_2014}. Although 
all results agree that soft particles move to the channel center, the influence of the Reynolds number is not completely clear. 
While in some cases the final equilibrium distance 
from the channel center
seems to depend on the Reynolds number\cite{kruger_interplay_2014, shin_inertial_2011}, 
\citet{kilimnik_inertial_2011} found no evidence of such a 
behavior in their computer simulations. In contrast, they showed
that the 
recorded values of the
equilibrium distance 
collapse
on a single master curve when plotted against the deformability of the particles. 

Using
additional 
control forces
along the channel axis
allows 
for
further adjustment of the 
particles'
equilibrium positions. In their 
experiments
\citet{kim_axisymmetric_2009} demonstrated that particles can be focused by applying an axial 
control
force. 
They used
an %
electric field 
to slow
down the particles relative to the fluid. Due to the induced Saffman force, the particles migrated towards the channel center\cite{saffman_lift_1965}. 
While these experiments were performed at small Reynolds numbers, previous simulation
work 
at moderate Reynolds numbers\cite{prohm_feedback_2014}
showed how an axial control force can move a rigid particle against the inertial
lift force to a new equilibrium position. 
This position
sensitively depends on the strength of the control force and the particle radius.
Finally, the inertial focusing of particles can also be influenced by controlling their rotational motion\cite{prohm_controlling_2014}
or by designing external force fields with the help of optimal control theory \cite{prohm_optimal_2013}.

In this 
article
we study the dynamics of a single deformable 
elastic
capsule 
flowing through a microchannel with quadratic cross section. To simulate the Newtonian fluid, we employ
the lattice-Boltzmann method together with the immersed boundary method. 
In the first part 
we study in detail
the equilibrium positions of the elastic capsules,
their deformations in the inertial Poiseuille flow,
as well as the scaling of the lift-force
profile by varying
Reynolds number, 
particle rigidity, and radius.
In an intermediate regime of particle rigidity, we
confirm that the equilibrium distance
hardly depends on Reynolds number and also identify
a similar 
master curve for varying
particle radius. The lift-force profiles
behave in a similar 
manner as
those for rigid particles. 

In the second part we apply an external axial force to
allow 
for an
additional control of the equilibrium position. 
Extending the study of
rigid particles
in Ref.\ \cite{prohm_feedback_2014},
we find that a 
speed-up of the particle
first
induces a drift 
further away from the channel center, which then reverses for even stronger speed-up.
This effect becomes
more pronounced as the particles 
are
softer. For very soft particles the behavior 
is
opposite 
to rigid particles.
A strong speed-up pushes them to the channel center while they move away from the center when slowed
down. Interestingly,
for a specific external control 
force
all particles assemble at the same equilibrium distance independent of their deformability. 

The article is organized as follows. In sect.\ \ref{sec.methods} we explain the set-up of our system, some details of the
lattice-Boltzmann method, how we model deformable capsules, and the measurement of the lift-force profiles.
In sect.\ \ref{sec.results} we present the results including the detailed study of the equilibrium position, the deformations of the
capsules, the lift-force profiles, and finally the influence of an external control force. We summarize and close with final
remarks in sect.\ \ref{sec.conclusions}.

\section{Methods} \label{sec.methods}

\subsection{Microfluidic setup in the simulations}

\begin{figure}
\includegraphics[width=.57\linewidth]{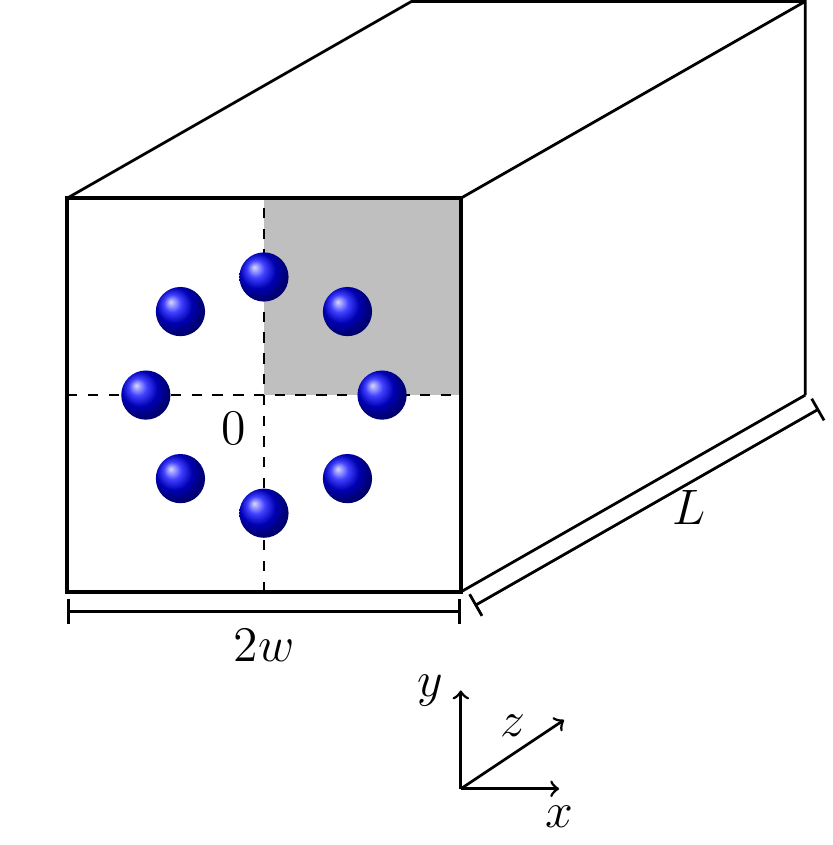}
\includegraphics[width=.35\linewidth]{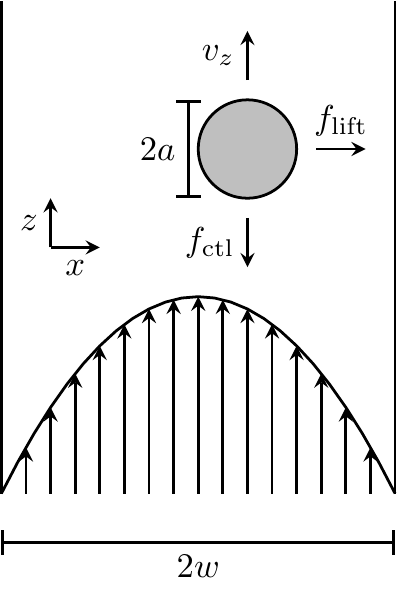}
\caption{Left: Schematic of the microfluidic setup in the simulations. 	We use a quadratic channel with edge length $2w$ and channel length $L$. The channel center is located at $x,y=0$. Due to the symmetry of the quadratic cross section, only the shaded area needs to be con\-si\-dered.	The blue spheres indicate possible stable equilibrium positions.
Right: Poiseuille flow in the $xz$ plane. The undeformed capsule with radius $a$, lift ($f_{\mathrm{lift}}$) and control ($f_{\mathrm{ctl}}$) forces acting on it, and the axial capsule velocity $v_z$ are indicated.}
\label{fig:setup}
\end{figure}

We study a single deformable elastic capsule flowing in a microfluidic 
channel with a quadratic cross section (edge length $2w$ and channel length $L$; see \prettyref{fig:setup}). The pressure-driven Poiseuille flow points along the $z$ direction and the origin of the coordinate system is in the channel center. The channel is filled with a Newtonian fluid with density $\rho$ and kinematic viscosity $\nu$.
A constant local body force (see sect.\ \ref{subsec.LB}) generates 
the
almost parabolic Poiseuille flow
of a square channel
\cite{bruus_theoretical_2008}.
It is characterized by the channel Reynolds number ${\re = 2w u_\text{max} /\nu}$, where $u_\text{max}$ is the maximal flow velocity in the channel center.

Inside the channel we place a deformable, neutrally buoyant particle,
which is filled with a fluid with the same viscosity as the surrounding medium. We model the deformable capsule by the Skalak model and characterize its deformability by the Laplace number (see sect.\ \ref{subsec.deform}). In such a setup the capsule shows cross-streamline migration within the cross section due to three effects:
i) Already at zero Reynolds number deformable capsules migrate towards the channel center, where shear rate vanishes, even in unbounded parabolic flow \cite{kaoui_lateral_2008}. This effect is important for the F\aa hr\ae us-effect\cite{fahraeus_suspension_1929}.
ii) At non-vanishing Reynolds number the parabolic flow profile generates a dynamic pressure difference along the particle surface.
Thereby
an inertial lift force occurs that drives the particle outwards towards the channel walls as indicated in \prettyref{fig:setup} \cite{segre_behaviour_1962}.
iii) Finally, a wall-induced lift force acts against the inertial lift force \cite{zeng_wall-induced_2005}.
All three effects ultimately determine the equilibrium position of the particle in the channel cross section.

The 
time
for migrating towards the equilibrium position
depends on the fluid parameters. Typically,
in our simulations
the particles assembled at their final 
location
after 10 to 100 $w^2/\nu$. 
With a channel width of \SI{50}{\mu m} and 
the kinematic
viscosity of
water,
\SI{1e-6}{m^2/s}, this corresponds to \SIrange{2.5e-2}{0.25}{s}.
Our longest simulations are equivalent to \SI{1}{s}.

In addition, an external control force applied on the capsule along the channel axis can tune this equilibrium position. For example, when directed against the flow  (see \prettyref{fig:setup}), the positive control force $\vec f_\text{ctrl}$ slows down the particle.
Thereby, the relative velocity between particle surface and surrounding fluid changes. This modifies the dynamic pressure difference along the particle surface and a lateral migration known as Saffman effect occurs \cite{saffman_lift_1965}.

\subsection{Lattice-Boltzmann method}
\label{subsec.LB}

\begin{figure}[bt]
\centering
\includegraphics[width=.6\linewidth]{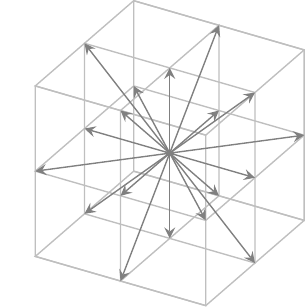}
\caption{Set of velocity vectors for the D3Q19  implementation of the LBM. Six vectors point to the center of the faces, 12 to the center of the edges, and one is the zero velocity.}
\label{fig:lattice}
\end{figure}

The viscous fluid is simulated by the lattice-Boltzmann method (LBM) in 3D with 19 different velocity vectors (D3Q19) using the Bhatnagar-Gross-Krook (BGK) collision model \cite{succi_lattice_2001,bhatnagar_model_1954}. The LBM 
generates a solution of
the Navier-Stokes equations by solving the Boltzmann equation on a cubic lattice. For this the one-particle distribution function $f(\vec x,\vec v,t)$  for position $\vec x$ and velocity $\vec v$ is discretized in space on a cubic lattice and for a discrete set of velocity vectors. These vectors point to the center of the faces and edges of a cube  (\prettyref{fig:lattice}). The time evolution of the Boltzmann equation is governed by two alternating steps: 
\begin{align}
\text{collision: } &f_i^*(\vec x,t) = f_i(\vec x,t)+\frac 1\tau \left[f_i^\text{eq}(\vec x,t)-f_i(\vec x,t)\right] \\
\text{streaming: } &f_i(\vec x+\vec c_i\Delta t,t+\Delta t) = f^*_i(\vec x,t) \, ,
\end{align}
where $f_i(\vec x,t)$ is the discretized particle distribution function with the index $i$ indicating the velocity vectors, $f_i^\text{eq}$ is an expansion of the Maxwell-Boltzmann distribution up to second order in the mean velocity, and $\tau$ is a typical relaxation time from the BGK model. 

Macroscopic quantities like the density $\rho$ or the momentum density $\rho\vec u$ correspond to the first two moments of the distribution function with respect to velocity:
\begin{align}
\rho(\vec x,t) &= \sum_i f_i(\vec x,t)\\
\rho(\vec x,t)\vec u(\vec x,t)&=\sum_i \vec c_i  f_{i}(\vec x,t) \, .
\end{align}
Following the Chapman-Enskog theory \cite{dunweg_lattice_2008}, %
one can derive a relation between the relaxation time $\tau$ and the fluid viscosity $\nu$,
\begin{equation}
\nu = c_s^2\Delta t \big(\tau-\frac{1}{2} \big)  \, ,
\label{eq.nu}
\end{equation}
where 
$c_s^2=\frac{1}{3}$
is the speed of sound measured in LBM units. It is important to note that the LBM is only valid for small Mach numbers $\ma=u_\text{max}/c_s$ to ensure the incompressibility of the Newtonian fluid. We chose the maximum flow velocity such that $\ma\le 0.1$ for all our simulations, which corresponds to density variations of less than 1\%. 

Simulations with immersed-boundaries like our deformable capsules show the best accuracy for relaxation times $\tau \le 1$ 
or viscosities $\nu\le 1/6$,
according to Eq.\ (\ref{eq.nu}) with LBM time step $\Delta t =1$.
We tuned $\re$ by varying $\nu$ through $\tau$ for Mach number $\ma=0.1$. When $\tau=1$ was reached,
we fixed this time and $\nu= 1/6$ and lowered $\ma$ to obtain the desired
Reynolds numbers
$\re < 50$.

To simulate the pressure-driven Poiseuille flow, we implemented a constant body force $\vec g$ following the Guo force scheme \cite{guo_discrete_2002}. 
In this scheme the
fluid momentum density at each time step is modified according to
\begin{equation}
\rho \vec u = \sum_i c_i f_i+\frac{\Delta t}{2}\vec g \, .
\end{equation}     
As lattice-Boltzmann solver we used the code provided by the Palabos project \cite{_palabos_2013}. The deformable capsule was implemented with the help of the immersed-boundary method as described in the following section.

Our simulations
were conducted 
with a length-to-width ratio of $L/2w = 4$ using periodic boundary conditions along the channel ($z$-axis).
As resolution for the lattice-Boltzmann solver, we chose 120 cubic unit cells along the full width of the channel, when we
recorded trajectories of the capsules, and 90 cells for determining the lift-force profiles. 
Especially for soft particles ($\la=1$ to 10) and high Reynolds numbers ($\re=100$) the lift force was not constant but oscillating in time.
To avoid these oscillations, the resolution was enhanced to 180 cells.

\subsection{Modeling a deformable capsule} 
\label{subsec.deform}

To couple the deformable capsule to the fluid, we use the immersed-boundary method 
\cite{peskin1972flow,peskin_immersed_2002,kruger_efficient_2011}.
In this method the capsule is modeled with a deformable Lagrangeian grid, which moves in the fixed Eulerian grid of the LBM. 
The 
Lagrangeian
mesh
of the undeformed spherical capsule
is obtained by 
starting from an icosahedron and then repeatedly subdividing each triangle into four triangles such that the distance
between two neighboring vertices is approximately equal to the lattice spacing of the Eulerian grid.
The membrane forces are 
calculated on the
Lagrangeian mesh and interpolated to the grid of the fluid. 
Likewise, the 
velocities
of the mesh vertices
are determined by an interpolation from
the fluid velocity
vectors
of the surrounding 
grid nodes. For both interpolations a weighting function $\phi(r)$ is used, which averages over all 
lattice nodes within
a distance $\Delta x$ from the center, where $\Delta x$ is 2.  For further details we refer to \cite{kruger_efficient_2011}.

To model the capsule, three contributions need to be considered: in-plane elasticity of the capsule membrane, its bending
stiffness,
and volume conservation due to the incompressible fluid interior.
Elastic deformations are described by the Skalak model, which originally was introduced for red blood cells \cite{skalak_strain_1973}. The elastic energy is written in the strain invariants $I_1$ and $I_2$ of the 
Green strain tensor
as defined in
appendix
\ref{subsec.cauchy},
\begin{equation}
E_s=\int\limits_{A_0} \dint A \left[ \frac{\kappa_s}{12}\left( I_1^2+2I_1-2I_2\right)+\frac{\kappa_a}{12}I_2^2\right] \, ,
\label{eq:skalak}
\end{equation}
where $\kappa_s$ is the shear modulus governing shear deformations and
$\kappa_a$ is the area modulus, which penalizes 
changes in the
area of the
mesh triangles relative to the initial 
value.
Bending stiffness
is needed to prevent buckling
of the capsules
and the formation of cusps.
To quantify bending elasticity, we take
the Helfrich functional, 
which
can be discretized 
on the triangular surface
mesh\cite{helfrich_elastic_1973,gompper_random_1996},
\begin{equation}
E_b = \sqrt{3}\kappa_b \sum_{ \langle ij\rangle} 
[1-\cos(\theta_{ij}-\theta_{ij}^{(0)})]
\,
\end{equation}
where $\kappa_b$ is the bending modulus, $\theta_{ij}$ is the angle  
between 
the normal vectors of two neighboring triangles, and $\theta_{ij}^{(0)}$ is the reference angle for the spherical capsule.
Finally, to approximately
ensure volume conservation, we added 
an energy
of the form 
\begin{equation}
E_v = \frac{\kappa_v}{2}\frac{ \big( V-V^{(0)} \big)^2}{V^{(0)}} \, ,
\end{equation}
where
$\kappa_v$ is the 
elastic bulk modulus,
$V$ the volume of the capsule,
and $V^{(0)}$ the volume of the spherical ground state.
We chose $\kappa_v$ such that the volume changes
were
less than 1\%.
To reduce the 
number of variable parameters,
we followed \citet{kruger_interplay_2014} 
and fixed the ratios of
the area and the bending 
moduli with the shear modulus,
\begin{equation}
\kappa_a/\kappa_s=2 \qquad \kappa_b/(\kappa_s a^2)=\num{2.87e-3}.
\end{equation}

The capillary number $\ca$, as the ratio between between applied viscous shear stress  $\rho\nu u_\text{max}/w$ and typical elastic stress $\kappa_S/a$, describes the deformability of an elastic capsule,
\begin{equation}
\ca = \frac{\rho \nu u_\text{max}a}{w\kappa_s} \, .
\end{equation}
The capillary number still depends on the explicit flow speed. A dimensionless number to quantify the rigidity of
a capsule is the Laplace number. It is defined as the ratio between typical elastic shear forces $\kappa_s a$ and the intrinsic
viscous force scale $\rho\nu^2$. Using the particle Reynolds number  $\re_{p} = \re (a/w)^2$, one can write the Laplace
number as
\begin{equation}
\la = \frac{\re_p}{\ca} = \frac{\kappa_s a}{\rho\nu^2}
\end{equation}
We will see that it is the right quantity to plot the equilibrium distance of a capsule
from the channel center.

\subsection{Measurement of the lift-force profile}
\label{sec:lift_force}
To 
determine
the inertial lift force acting on the capsule, we fix its lateral position and measure the 
force 
of the LB
fluid 
on the membrane
following\cite{prohm_controlling_2014}.
To hold the 
capsule
in place, we apply 
an adjustable force $f_\text{fb}$ evenly distributed over all the membrane vertices,
which compensates the 
lift force.
To calculate this feedback force, we use a proportional-integral (PI) feedback controller
\cite{astrom_feedback_2008}.
The idea is to steer a system parameter, here the lateral position, by an external force $f_\text{fb}$ in a targeted state.
In our case the
dynamics of the lateral position 
of the capsule's center of mass, $\vec x_\text{com}$,
is determined by
\begin{equation}
\dot{\vec x}_{\text{com}} = \vec f_\text{lift}+ \vec f_\text{fb} \, . %
\end{equation}
The feedback force depends on the 
deviation of the control parameter from
the desired state
at the current and all previous times,
\begin{equation}
\vec f_\text{fb}(t)=\gamma_x [\vec x_0-\vec x_\text{com}(t) ] + \int\limits_0^t \gamma_u [\vec x_0-\vec x_\text{com}(t^\prime)]
\dint t^\prime \, ,
\end{equation}
where $\gamma_{u/x}$ 
are
the stiffnesses of the feedback controller
and $\vec x_0$ 
is the targeted 
lateral
position.
In 
steady state, 
$\dot{\vec x}_{\text{com}} = 0$, one obtains
$\vec f_\text{lift}= -\vec f_\text{fb}$. The lift force is averaged over at least 200000
time steps, where the first 20000 time steps, until 
the system equilibrates,
are neglected.

\section{Results} \label{sec.results}

\subsection{Equilibrium position} \label{sec:equi_pos}

We first study in detail the equilibrium position of one single capsule varying particle radius $a$, 
its
deformability 
measured by the 
capillary number $\ca$ or the
Laplace number $\la$, and the channel Reynolds number $\re$.
We vary the Laplace number in the interval between 1 (very soft) and 1000 (almost rigid)
or the capillary number between $10^{-4}$ and 10. For the Reynolds number
we choose the values \numlist{5;10;50;100}.

\begin{figure}%
\includegraphics[width=\linewidth]{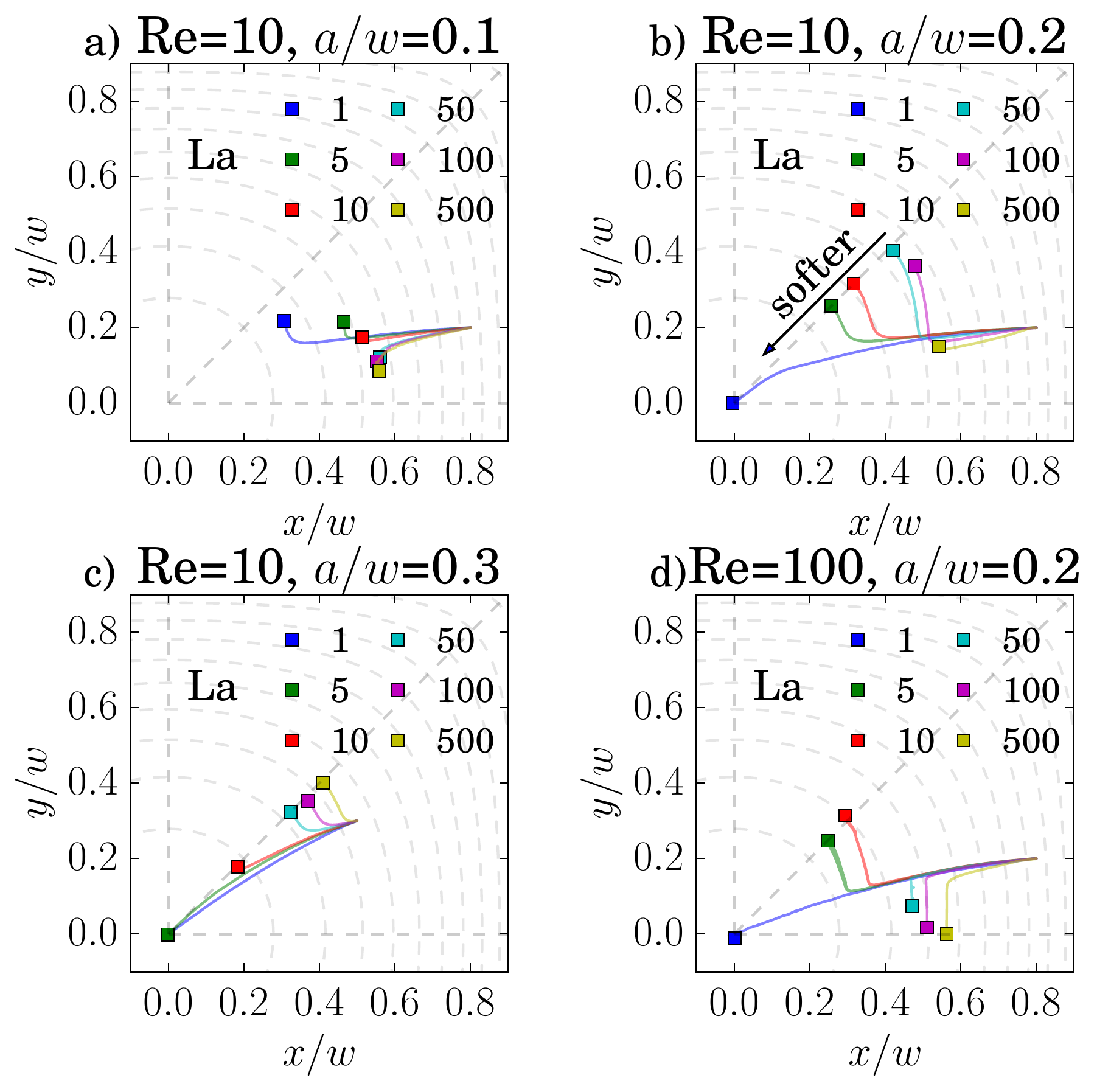}
\caption{Trajectories of the capsules in the cross-sectional plane of the channel with different rigidities quantified by the 
Laplace number $\la$. The trajectories are plotted for different particle sizes $a/w$ and Reynolds numbers $\re$ as indicated 
in a)-d). The capsules start at the same position and the endpoint of the trajectories are shown by filled squares. Not all of 
them reach their equilibrium position on the diagonal or the main axis during the simulations.}
\label{fig:traj_scatter}
\end{figure}

\begin{figure}%
\center
\includegraphics[width=0.9\linewidth]{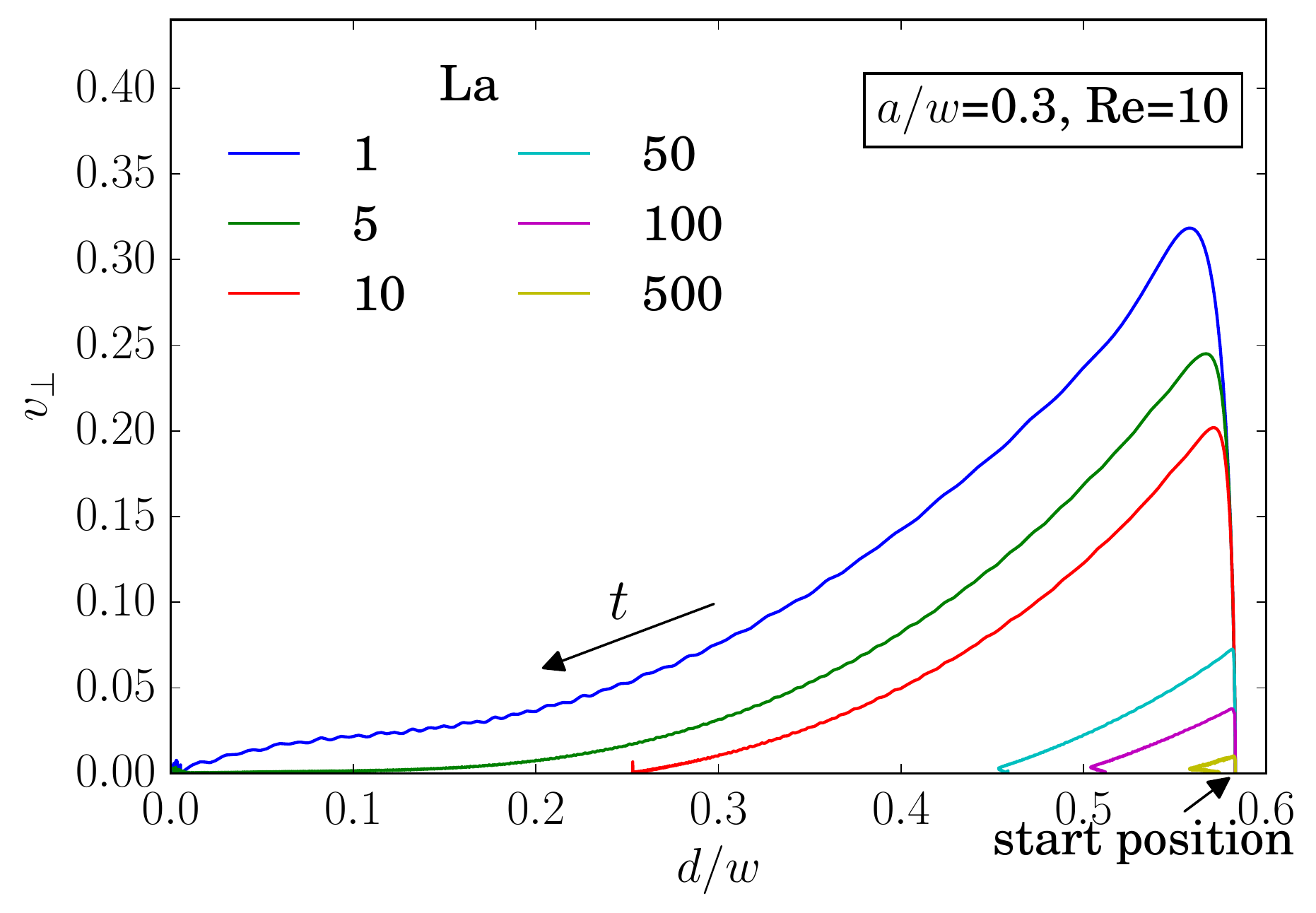}
\caption{ Lateral velocity $v_{\perp}$ plotted versus the distance $d$ to the channel center	for different Laplace numbers $\la$. 	All particles start at the same initial position ($d/w=0.583$) and with time move inwards to their equilibrium positions. 
}
\label{fig:lateral_velocity}
\end{figure}

We used two different methods to determine the equilibrium position of the capsule. The first method is closer to the experiment. The particle is put at a specific position and can then move freely. 
Figure\ \ref{fig:traj_scatter} shows a collection of example trajectories. Especially the small particles with $a/w = 0.1$ 
at $\re = 10$ migrate slowly and do not reach their equilibrium positions on either the diagonal or the $x$ axis, even in the 
longest simulation runs. 
The reason is that the lift force driving the particle motion strongly scales with particle radius $a$ and Reynolds number, 
as discussed in sect.\ \ref{subsec.lift} but also depends on the deformability of the capsule. However, in most cases one 
clearly sees where the particles migrate to.

For the particle trajectories, which we visualize in  \prettyref{fig:traj_scatter}c), we plot in \prettyref{fig:lateral_velocity} the lateral velocity $v_{\perp}$ in the cross-sectional plane of the channel versus the distance $d$ from the center.
All particles start at the same position at a distance $d/w = 0.583$. 
The velocity rapidly increases and then gradually decreases to zero as the particles move inwards towards their equilibrium position.
Clearly, softer particles (small $\la$) show a larger lateral migration velocity than rigid particles. 
For the smaller particles in \prettyref{fig:traj_scatter}b) the migration process itself splits into two phases. First,
the particles quickly move inwards along the radial direction. This corresponds to the gradual decrease of the lateral velocity as 
discussed in \prettyref{fig:lateral_velocity}. When the capsules are close to their final equilibrium distance, the radial movement 
stops and the lateral velocity is close to zero. Thus, the particles only very slowly drift along the polar direction towards their 
equilibrium positions. A similar kind of motion was already observed for rigid particles \cite{chun_inertial_2006}.

\begin{figure}%
\center
\includegraphics[width=0.95\linewidth]{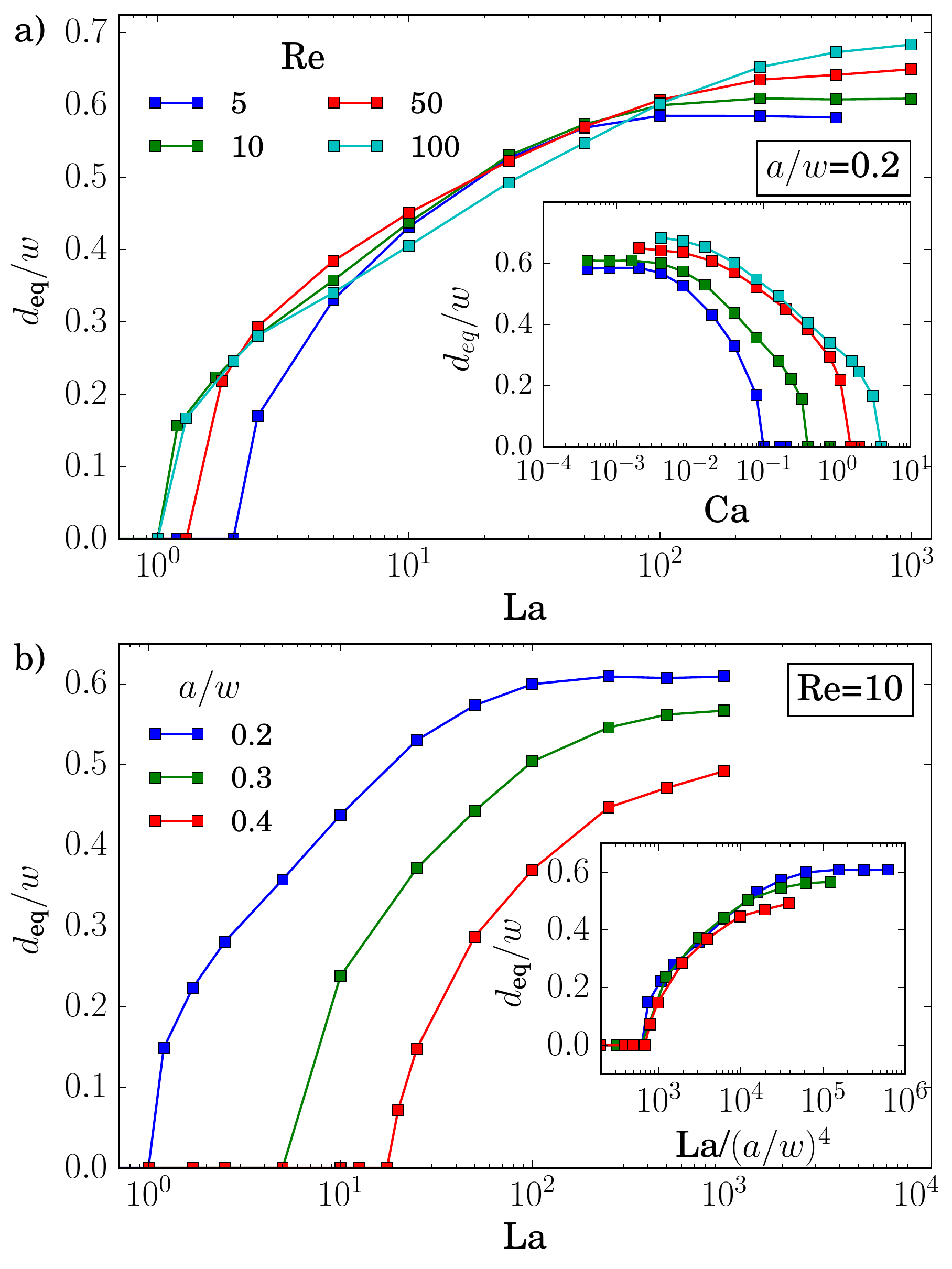}
\caption{
Equilibrium distance $d_{eq} /w$ from the channel center:
a) plotted versus Laplace number $\la$
for $a/w = 0.2$ and different $\re$,
Inset: $d_{eq} /w$ versus capillary number $\ca$;
b) $d_{eq} /w$ plotted versus $\la$ 
for $\re=10$ and different $a/w$,
Inset: $d_{eq} /w$ versus $\la / (a/w)^4$.
}
\label{fig:equi_pos}
\end{figure}

To have a computationally more efficient method for determining the equilibrium positions, we also measured 
lift-force profiles using the method outlined in \prettyref{sec:lift_force}. 
The equilibrium positions correspond to the stable fix points. We will discuss the lift-force profiles in more detail in sect.\ \ref{subsec.lift}.
Both methods agree well, as we demonstrate in \prettyref{fig:compare_equi_pos} in appendix\ \ref{subsec.method}. For the following discussion 
based on \prettyref{fig:equi_pos} we will take the equilibrium positions determined from the lift-force profiles. The free migration of 
the  capsules indicate that 
they mostly lie on the diagonals. Only 
small capsules ($a/w=0.1$) and 
capsules at $\re=100$, which are sufficiently rigid  ($\la>50$), clearly migrate towards the $x$ axis.
The results on the rigid particles are in agreement with our earlier work\ \cite{prohm_feedback_2014}.

\begin{figure}
\includegraphics[width= 0.9\linewidth]{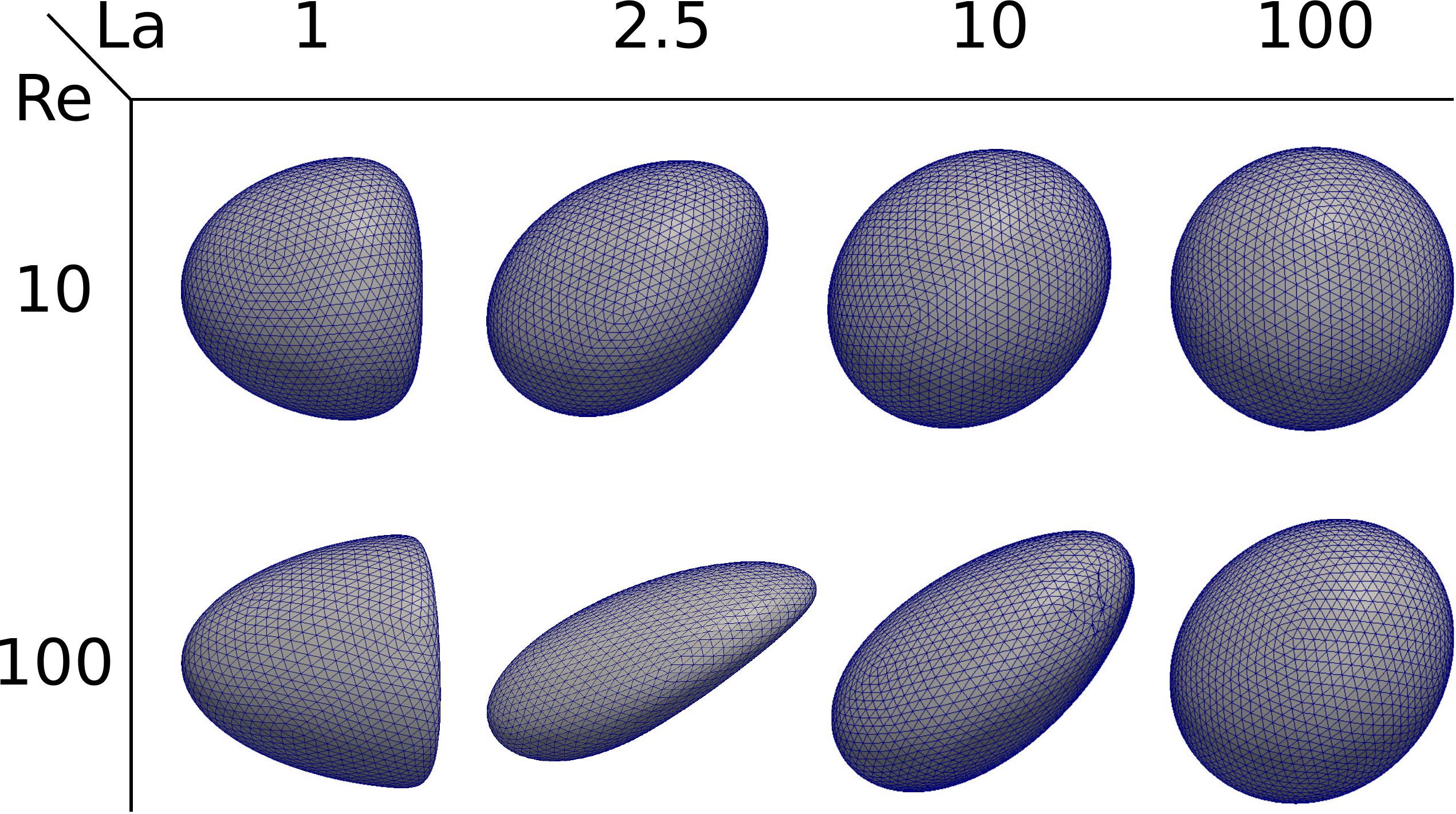}
\caption{%
Shapes of the capsules for different Laplace and Reynolds numbers.}
\label{fig:deformation_capsules}
\end{figure}

We now discuss the equilibrium distance from the channel center, $d_{eq}$, in  more detail. In the inset of 
Fig.\ \ref{fig:equi_pos}a) we plot it versus the deformability of the capsule, the capillary number $\ca$, for different Reynolds 
numbers $\re$. All curves show how deformable capsules (large $\ca$) migrate to the channel center. Making them 
more rigid by decreasing $\ca$, the inertial lift force becomes more dominant and drives the capsules away from the center
towards the equilibrium distance of solid spheres.
Clearly, since the strength of the inertial forces increases with $\re$, the migration towards the channel walls already starts 
when the capsules are more deformable (large $\ca$).

All curves in the inset roughly have the same shape and shift to the
right with inreasing $\re$. The capillary number $\ca$ is proportional to the absolute flow velocity $u_\text{max}$. 
Indeed, removing this dependence and plotting the equilibrium distance versus the Laplace number $\la$, which measures
the rigidity of the capsule, all curves roughly fall onto one master curve [see main plot of Fig.\ \ref{fig:equi_pos}a)]. This was 
already observed by \citet{kilimnik_inertial_2011}. Deviations from the master curve occur for very rigid capsules (large $\la$),
which move closer to the channel walls with increasing $\re$ \cite{asmolov_inertial_1999, matas_inertial_2004,prohm_inertial_2012}.
Also, the value of $\la$, where the capsule starts to move away from the channel center, is sensitive to $\re$.
This might be due to the fact that the shapes of two capsules, located either in the channel center or close-by, differ strongly
as we will discuss below.

Also the particle radius $a/w$ plays an important role for the equilibrium distance, as Fig.\ \ref{fig:equi_pos}b) demonstrates.
Larger particles leave the channel center at larger $\la$ and thereby assemble closer to the channel center compared to smaller 
particles. One obvious reason for this behavior is that $\la \propto a$. However, in addition the lift force, which drives the
capsule away from the center, roughly scales with $a^3$ as discussed in \cite{di_carlo_particle_2009} and Sec.\ \ref{subsec.lift}. 
So, if we plot $d_{eq}$ versus $La / (a/w)^4$, the equilibrium distances collapse on a single master curve
in the regime where the capsules are deformable [see inset of Fig.\ \ref{fig:equi_pos}b)].
For rigid particles (large $\la$) the curves do not collapse since smaller
particles move closer to the channel walls \cite{prohm_inertial_2012,hur_deformability-based_2011}.

\subsection{Deformation of the capsules} \label{sec:deform}

\begin{figure}
\includegraphics[width=\linewidth]{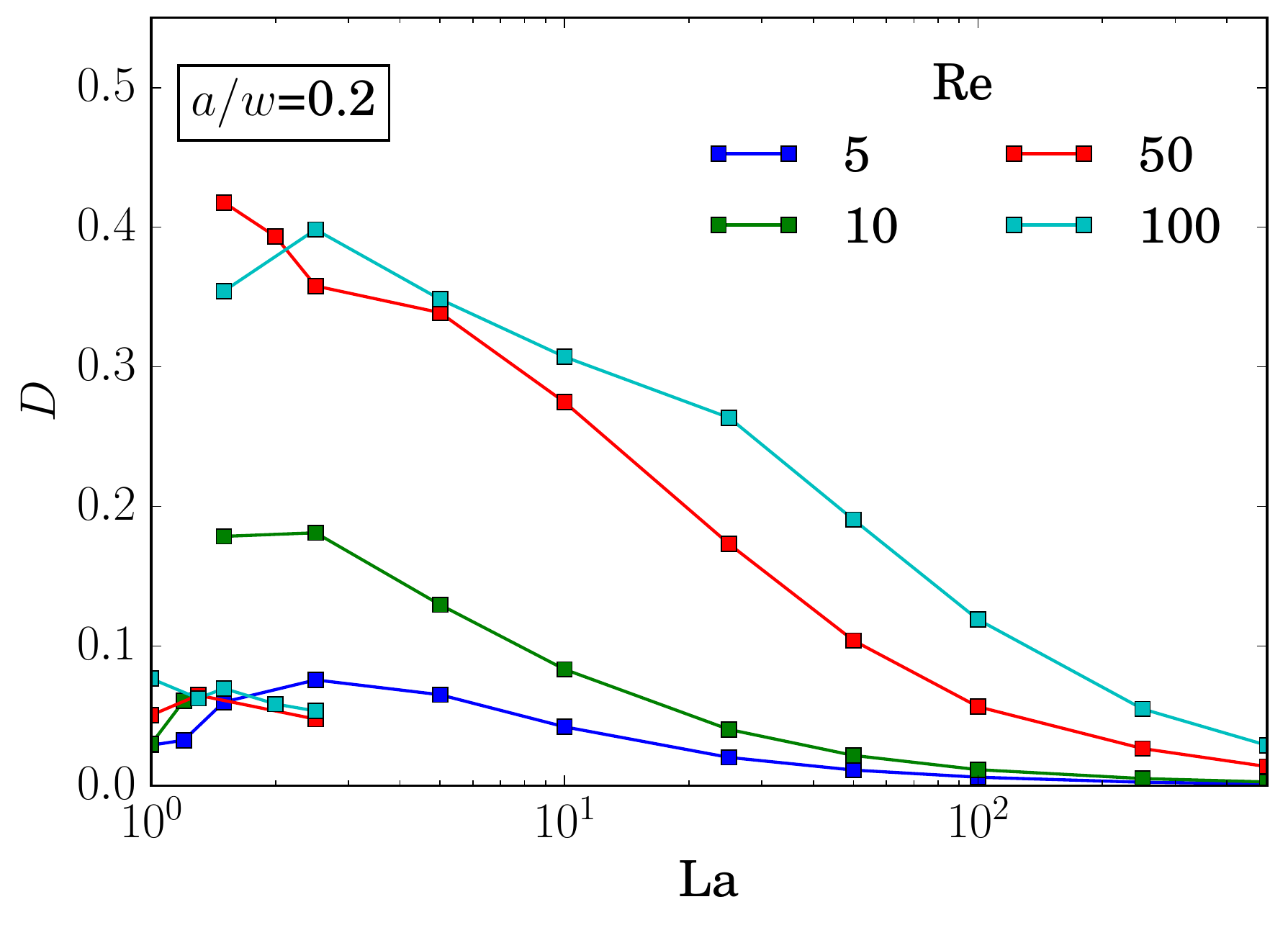}
\caption{Taylor deformation index $D$ of 
an elastic capsule at the equilibrium location
plotted versus Laplace number $\la$ for different Reynolds numbers $\re$. 
}
\label{fig:deformation}
\end{figure}

In Fig.\ \ref{fig:deformation_capsules} we illustrate the shapes of capsules at their equilibrium positions for different 
Laplace ($\la$) and Reynolds ($\re$) numbers. At $\la = 1$ both capsules are located in the channel center and show the
expected form of a parachute, which is more visible at larger Reynolds numbers\cite{horwitz_three-dimensional_2014}. At $\la=2.5$, the capsules have left the center 
with $d_{eq}/w \approx 0.28$ and the deformation is more asymmetric. The capsules become less deformed with increasing 
$\la$, although the capsules move further away from the channel center, where the viscous shear stresses increase. 
Interestingly, although the shapes of capsules with the same $\la$ differ for the two $\re$ values, in Fig.\ \ref{fig:equi_pos}a) we 
demonstrated that their equilibrium distances to the channel center are roughly the same independent of $\re$.

To quantify the deformation of the capsule, we follow 
\cite{taylor_formation_1934, ramanujan_deformation_1998}
and determine its moment-of-inertia tensor and 
then
the equivalent 
ellipsoidal particle with the same principal moments of inertia. 
Now, the
smallest and largest semi-axes
of the ellipsoid,
$r_\text{min}$ and $r_\text{max}$, 
are then used to define the Taylor deformation index,
\begin{equation}
D=\frac{r_\text{max}-r_\text{min}}{r_\text{max}+r_\text{min}} \, .
\end{equation}
For spherical particles it is zero.
In Fig.\ \ref{fig:deformation} we plot $D$ versus Laplace number $\la$ for different $\re$.
The parachute form of soft particles ($\la$=1) at the channel center gives a small $D$. 
For $\re=5$ when
the capsules move away from the center, the deformation index 
increases 
smoothly
and goes through a maximum, which is approximately situated at $\la \approx 2.5$.
Instead, for $\re=10$, 50, and 100 we observe an abrupt increase of $D$ to another branch.
At this point the equilibrium position in the center becomes unstable with increasing $\la$ and sharply increases
from zero, as visible in Fig.\ \ref{fig:deformation_capsules}a). However, in our simulations
we hardly see the capsules moving away from the center within a simulation time of several million time steps. This 
is indicated by the extended branches at small $D$. For
all $\re$ values
we find increasing
$\la$ further also reduces $D$. 
For $\re=100$ a kink is visible at $\la \approx 25$. Here the capsule switches its equilibrium location from the diagonal of
the channel cross section to one of the main axes.

\subsection{Lift-force profile} 
\label{subsec.lift}

\begin{figure}[bt]
\includegraphics[width=\linewidth]{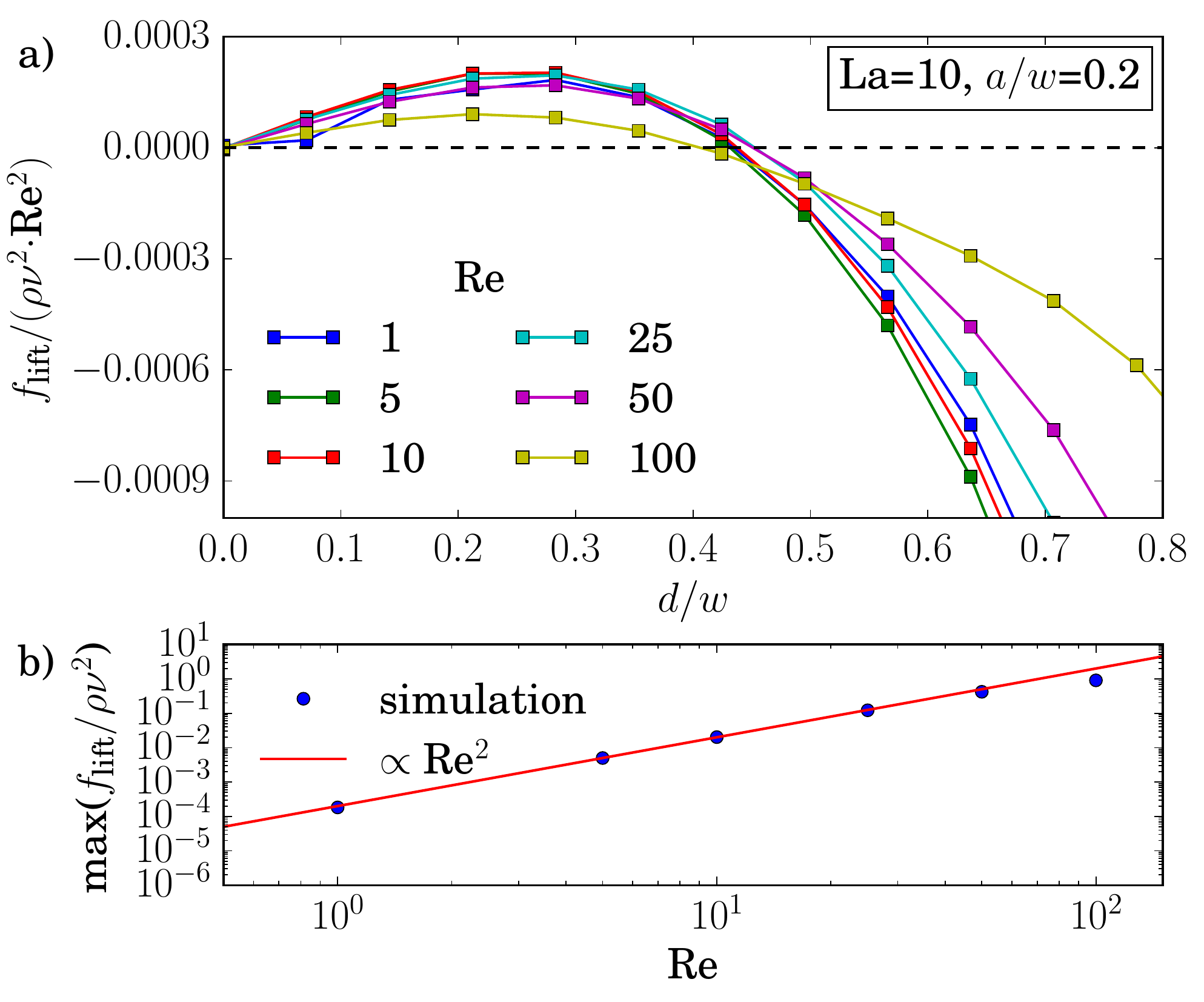}
\caption{
a) Inertial lift force $f_{\mathrm{lift}}$ along the diagonal in units of $\rho \nu^2 \re^2$ plotted versus distance 
$d/w$ to the center
for different Rey\-nolds numbers. 
Other parameters are $\la=10$ and $a/w=0.2$.
b) Maximum of $f_{\mathrm{lift}}$ plotted versus $\re$.
}
\label{fig:lift_force_reynolds}
\end{figure}
\begin{figure}[bt]
\includegraphics[width=\linewidth]{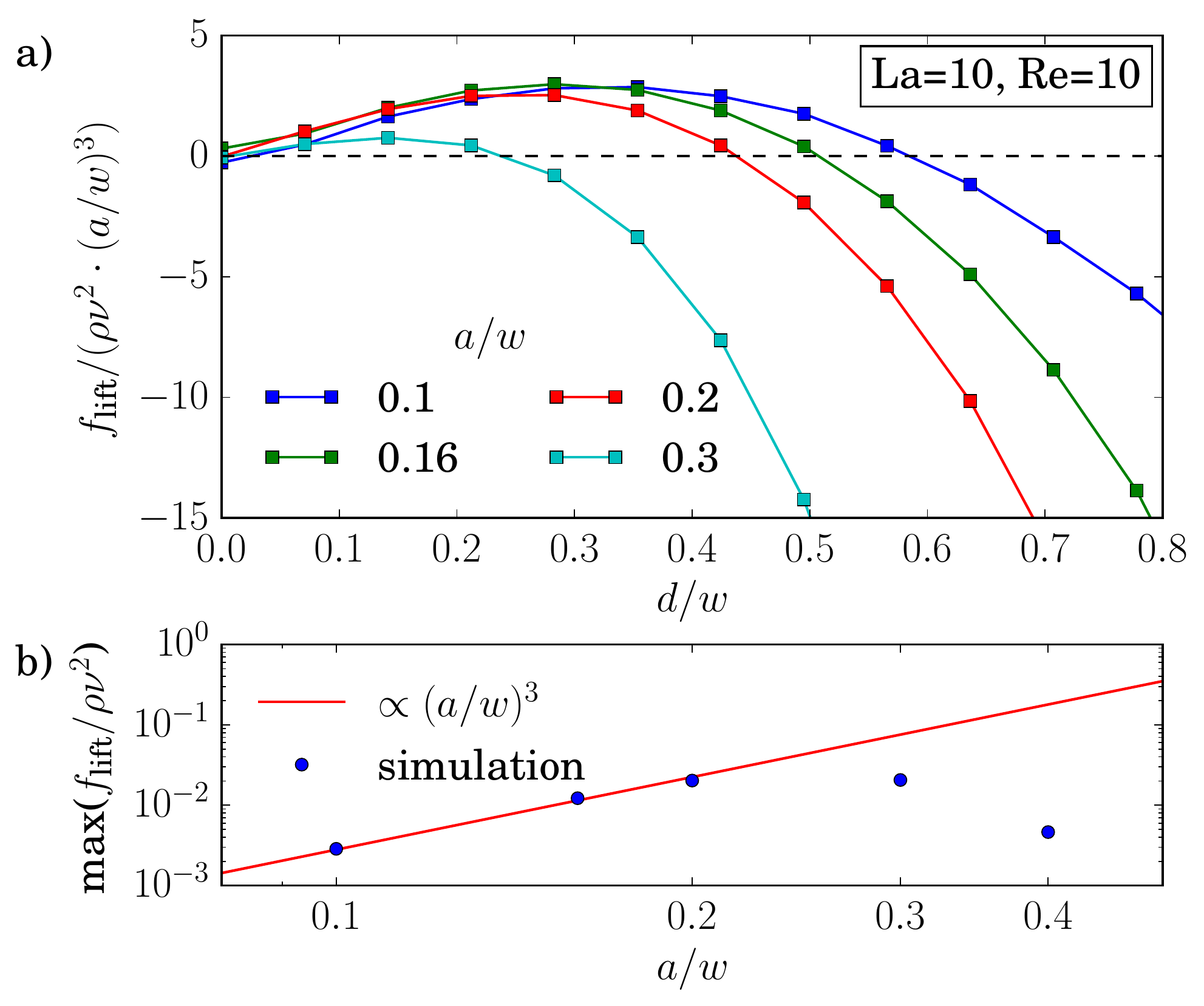}
\caption{
a) Inertial lift force $f_{\mathrm{lift}}$ along the diagonal in units of $\rho \nu^2 (a/w)^3$ plotted versus distance 
$d/w$ to the center
for different particle radii $a/w$.
Other parameters are $\la=10$ and $\re=10$.
b) Maximum of $f_{\mathrm{lift}}$ plotted versus $a/w$.
}
\label{fig:lift_force_radius}
\end{figure}
\begin{figure}[bt]
\includegraphics[width=\linewidth]{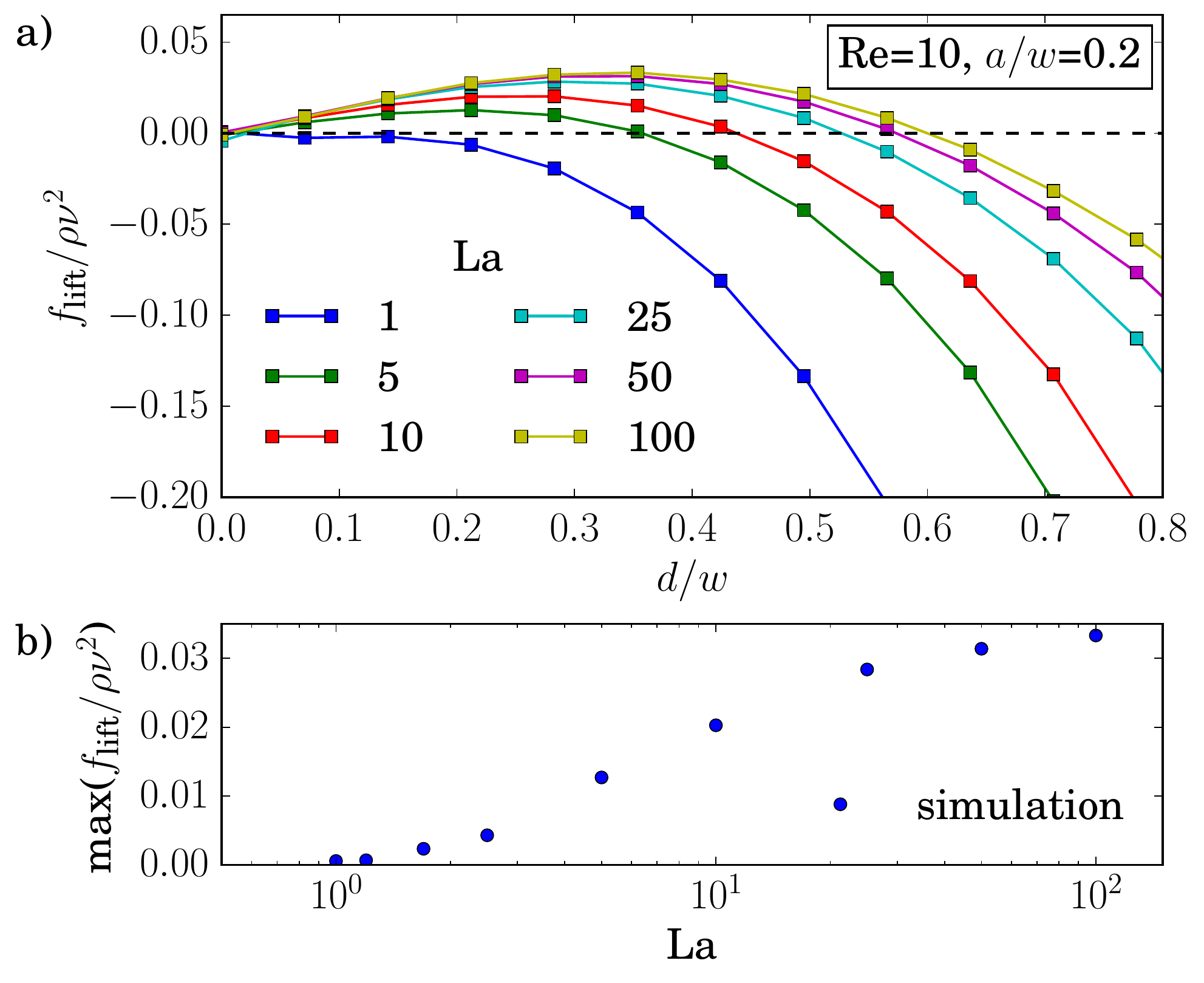}
\caption{
a) Inertial lift force $f_{\mathrm{lift}}$ along the diagonal in units of $\rho \nu^2$ plotted versus distance 
$d/w$ to the center
for different Laplace numbers. 
Other parameters are $\re=10$ and $a/w=0.2$.
b) Maximum of $f_{\mathrm{lift}}$ plotted versus
$\la$.
}
\label{fig:lift_force_laplace}
\end{figure}

The dynamics and equilibrium positions of an elastic
capsule are determined by the lift-force profile. In this section we study 
how the inertial lift force scales with
Reynolds and Laplace number
as well as particle radius
and compare it to rigid particles. 
For 
rigid particles
with radii much smaller than the channel width
the analytical solution\cite{segre_behaviour_1962,asmolov_inertial_1999} for the lift force predicts 
the scaling
$f_\text{lift} \propto \re^2(a/w)^4$. For larger rigid particles, of similar size 
as
ours, 
numerical solutions of the Navier-Stokes equations give a
scaling 
proportional to $\re^2(a/w)^3$ near the channel center and $\re^2 (a/w)^6$ near the 
channel
walls\cite{di_carlo_inertial_2009}.
To measure the lift force we use the procedure described in \prettyref{sec:lift_force}. 
As almost all capsules travel to equilibrium positions on the diagonal, we study the lift-force profiles along 
this direction.

We usually observe the
typical form of the lift-force profile with
two equilibrium positions of the capsule, where the lift force vanishes [see, for example, Fig.\ \ref{fig:lift_force_reynolds}a)]:
the unstable fix point in the channel center ($d=0$) and one stable fix point between channel center and walls.
In the previous section we found that the equilibrium position is almost independent of the Reynolds number, 
in particular, 
in an intermediate range of capsule rigidity measured by $\la$.
When we measure the lift-force profile for different Reynolds numbers 
while keeping the particle radius and the Laplace number fixed, 
this is also visible in the lift-force profiles as all zero crossings coincide [see \prettyref{fig:lift_force_reynolds}a)]. 
Furthermore, the profiles for small $\re$ fall on top of each other when we scale $f_{\mathrm{lift}}$ by $\re^2$. This
scaling is confirmed  in \prettyref{fig:lift_force_reynolds}b), where we plot the maximum
lift force 
versus $\re$.
Only for high Reynolds number ($\re=100$) we 
obtain
a deviation from $f_{\mathrm{lift}} \propto \re^2$.
Indeed, by measuring the particle-wall interaction of rigid particles, the authors of Ref. \cite{zeng_wall-induced_2005}
noticed an increase of the wall lift coefficient around $\re\approx 100$ due to an imperfect bifurcation of the wake structure.
This might explain our observation.
However, in total the scaling law $f_{\mathrm{lift}} \propto \re^2$ also seems to be valid for soft spheres 
at
moderate $\re$.

Figure\ \ref{fig:lift_force_radius}a) confirms the scaling $f_{\mathrm{lift}} \propto (a/w)^3$ for small distances,
while closer to the walls the force profiles clearly differ. Strong deviations also occur for larger particles with $a/w \ge 0.3$,
which is also visible in  the maximum lift force plotted versus $a/w$ in \prettyref{fig:lift_force_radius}b) . This is in contrast to
Refs. \cite{prohm_inertial_2012, di_carlo_particle_2009}, where the scaling was verified for particles with radii up to $a/w = 0.38$.
We
suspect
the different behavior to be due to the deformability of the capsules.

For increasing Laplace number the lift-force profiles approach the limit of a rigid capsule [see \prettyref{fig:lift_force_laplace}a)].
Making the particles softer, the stable equilibrium position moves
towards the channel center
and ultimately for $\la=1$ reaches the center, as already discussed in \prettyref{sec:equi_pos}.
Also, the 
maximum
value of the lift force decreases the softer the particles 
become
as illustrated in Fig.\ \ref{fig:lift_force_laplace}b).

\subsection{External control force}
\begin{figure}[bt]
\includegraphics[width=\linewidth]{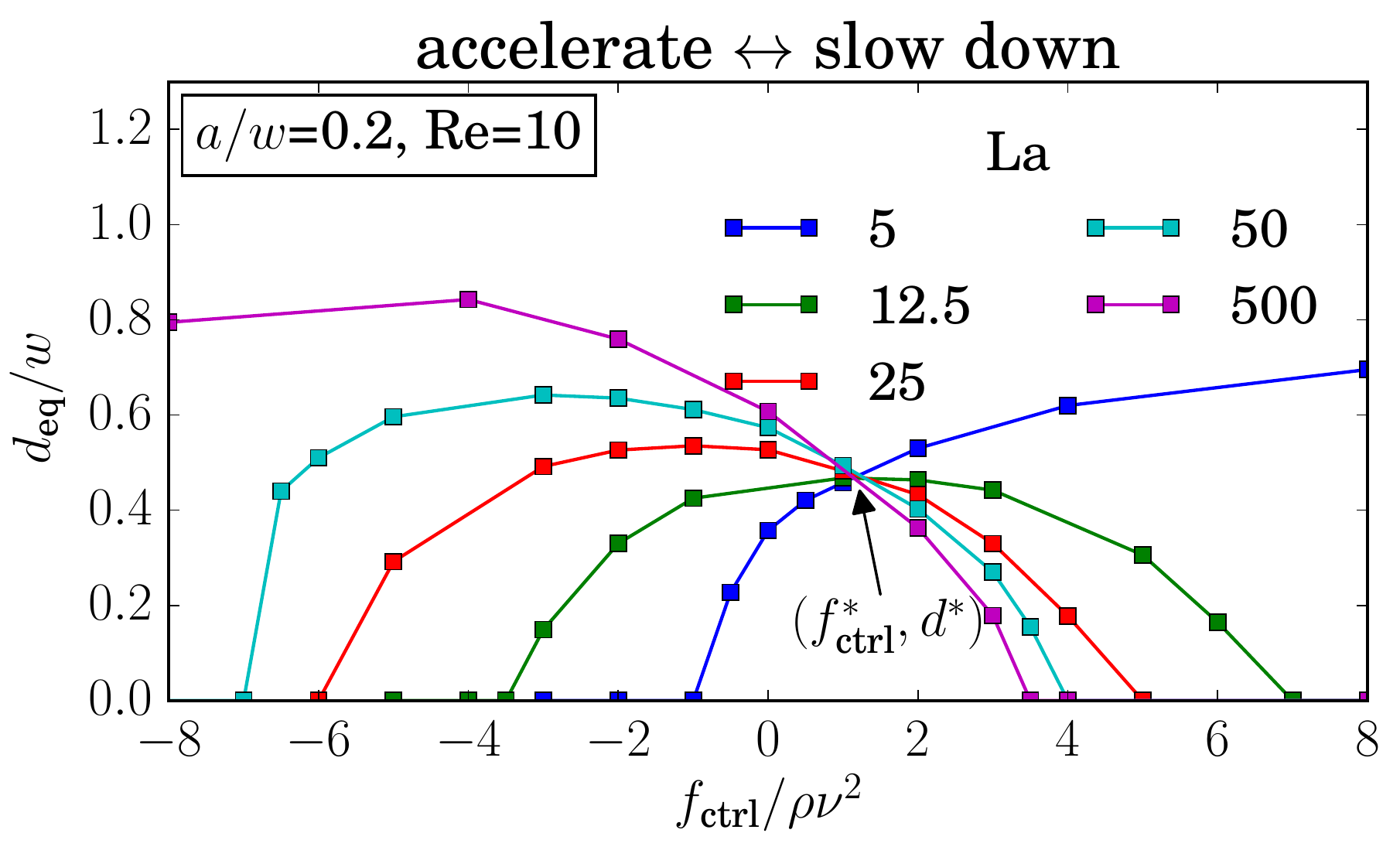}
\caption{
Equilibrium distance $d_{eq}$ 
plotted versus
the external control force $f_\text{ctrl}$ for different 
Laplace numbers $\la$. Other parameters are $\re=10$ and $a/w=0.2$.
}
\label{fig:control_diagram}
\end{figure}

Experiments\cite{kim_axisymmetric_2009} and simulations\cite{prohm_feedback_2014} showed that the 
lateral position of rigid particles can be controlled by an
external 
force,
applied along the channel axis. Depending on its direction it
either speeds up or slows down the particle relative to the channel flow.
This changes the relative velocity between particle surface and surrounding fluid. Thereby, 
the
dynamic pressure 
varies
along the surface, which causes the lateral motion of the particle known as 
Saffman effect\cite{saffman_lift_1965}.
In our simulations
we apply a constant force
on the elastic capsule evenly distributed
on the membrane vertices 
along
the axial direction. %
We choose the convention that
a positive force is directed against the flow 
and thus slows down the particle.
In \prettyref{fig:control_diagram} we plot
the equilibrium distance
to the channel center
as a function of the external control force $f_\text{ctrl}$.
As before,
we 
determined
the equilibrium positions from the stable fix points of the lift-force profiles.
For 
rigid 
capsules
($\la=500$) %
the results agree with previous simulations\cite{prohm_feedback_2014}: 
when these particles
are slowed down 
($f_\text{ctrl} > 0$),
they %
move towards the channel center and ultimately reach it for sufficiently large $f_\text{ctrl} > 0$,
while accelerated particles
move closer to the channel walls. 
However,
for strongly
negative control forces a decrease in the equilibrium distance is observable.
This behavior is clearer visible for $\la = 50$.
We observe
the softer the 
capsules are,
the more 
does the maximum equilibrium distance 
move to larger and even positive control forces. Furthermore, the maximum equilibrium distance decreases until
at around $\la \approx12.5$, where the curve is symmetric about its maximum.
For even softer 
capsules
it increases again.
As a consequence, soft particles ($\la=5$) behave
opposite to
rigid particles: when slowed down, they move away from the channel center, 
while when moving faster ($f_\text{ctrl} < 0$), they approach the center and ultimately stay there.

\begin{figure}[bt]
	\includegraphics[width=\linewidth]{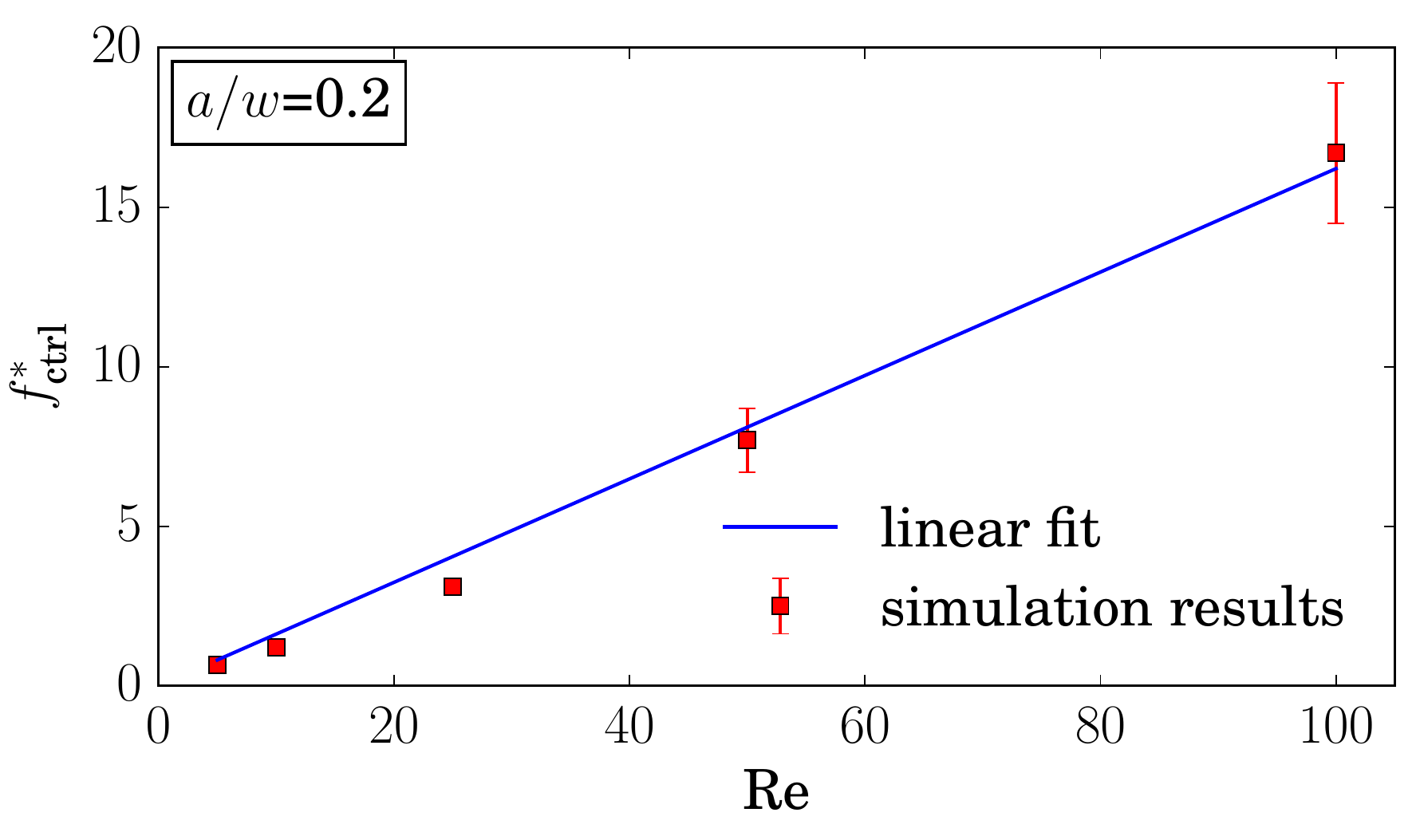}
	\caption{
	Control force of the intersection point
	$f_\text{ctrl}^\ast$ plotted versus 
        Reynolds number
        $\re$.
        At
         lower Reynolds numbers 
         the errorbars are 
         on
         the order of 
         the
         symbol size.        
         }
	\label{fig:control_intersection}
\end{figure}

Even more interesting, we find that all curves 
for different $\la$
intersect in one point 
at
a positive control force 
of about
$f_\text{ctrl}^\ast=1.2\rho\nu^2$.
At this control force all particles assemble at the same 
equilibrium distance $d_{\mathrm{eq}}^*$
independent of 
their
deformability. 
This property seems to be generic. We also find it for other
Reynolds numbers
as illustrated in \prettyref{fig:control_reynolds} of 
appendix\ \ref{sec.external}.
While
the intersection point moves to higher control forces with increasing Reynolds number,
in fact, $f_\text{ctrl}^\ast \propto \re $ as \prettyref{fig:control_intersection} demonstrates,
the distance from the channel center stays almost the same 
at
$\tilde d/w\approx0.46$.
We do not have a clear understanding of this behavior. However, we checked that it is generic and not a numerical artifact.
In particular, this behavior does not change when we increase the channel length to investigate the influence of
the periodic boundary condition or when we increase the resolution of the 
lattice-Boltzmann grid
to check for discretization errors
(see \prettyref{fig:check_control} in 
appendix\ \ref{sec.external}).

Using
such an external force, allows a much finer control of the particle's equilibrium position. For example, 
in the absence of a control force ($f_\text{ctrl}= 0$ in Fig.\ \ref{fig:control_diagram})
the equilibrium position of particles with $\la=5$ and $\la=12.5$ are quite close to each other.
By applying $f_\text{ctrl}=-1\rho\nu^2$ the softer particle moves towards the center while the equilibrium position of the less 
deformable particle hardly changes. This 
effect should 
help to
enhance particle separation based on its deformability.

\section{Conclusions}
\label{sec.conclusions}

Manipulating deformable capsules by hydrodynamic flow in the
regime of intermediate Reynolds numbers 
has a large variety of applications, e.g., for 
cell sorting or probing. The high fluid velocities allow a high throughput and the resulting inertial focusing can be used to separate 
and steer particles towards desired positions
within a channel.
In this work we 
studied the equilibrium 
positions of soft capsules while immersed in a Poiseuille flow through a quadratic channel. We find that most particles 
assemble along the diagonal of the channel. The softer the particles are the closer they
move
towards the channel center. Their final equilibrium distance depends on the particle deformability, its radius and the channel Reynolds number. By introducing the Laplace number as ratio of elastic 
force
to the intrinsic viscous force scale,
the equilibrium position for different Reynolds numbers 
collapse on a single master curve. The equilibrium distance is thus independent of the flow velocity. 
Additionally, we also identify such a date collapse
for different particle radii.

In contrast to the equilibrium distance we find that the deformation of the capsules strongly depends on the Reynolds number. The deformation of the off-centered capsules increases with decreasing Laplace number although the capsules migrate in areas with 
a smaller shear gradient. Very soft capsules assemble at the channel center and have a symmetric parachute shape.

The
lift-force profiles 
of
deformable capsules behave pretty much the same as 
those of
rigid particles,
where
the lift force scales 
as
$f_\text{lift}\propto \re^2(a/w)^3$. 
For deformable capsules, we
find deviations 
from
this scaling law only for high Reynolds number ($\re=100$) 
and large particles ($a/w=0.3$). We were not able to 
identify
a similar scaling 
involving
the Laplace number. For 
decreasing
$\la$, the stable 
equilibrium position moves
towards the channel center and the maximum value of the lift force decreases until only a stable fixpoint in the channel 
center 
remains.
Finally, we
demonstrated that the particle equilibrium position can be controlled 
by an
external 
force along the channel axis.
For almost rigid particles ($\la=500$) we confirmed previous results\cite{prohm_feedback_2014},
but 
found a new behavior for 
soft
particles. While rigid particles migrate towards the channel center when they are slowed down, 
moderately
soft particles 
with $\la\approx 12.5$
migrate towards the channel center 
both 
for
positive and negative control forces. For even lower $\la$ the capsules behave opposite 
to rigid particles as they move towards the channel wall when slowed down. 
Interestingly,
we observe that all graphs 
with
different Laplace numbers 
intersect
in one point 
at
a non-zero control force.

Cancer cells are softer than healthy cells \cite{suresh_biomechanics_2007}. 
Using an external control force, 
enhances the sensitivity to separate them based on the lateral locations they assume in microchannels.
Thus, our
new theoretical insights might prove useful in 
developing
new 
biomedical 
devices to probe and separate cells based on their size and deformability.

\section{Acknowledgments}
We thank C. Prohm for providing the source code and for initial work in the project.
We acknowledge support from the Deutsche Forschungsgemeinschaft in the
framework of the Collaborative Research Center SFB 910.

{\footnotesize{
	\bibliography{literature} %
	\bibliographystyle{rsc} %
}}

\clearpage

\appendix
\setcounter{figure}{0}
\renewcommand{\thefigure}{A\arabic{figure}} %
\setcounter{equation}{0}
\renewcommand{\theequation}{A\arabic{equation}}
\section{Appendices}

\subsection{Determining the equilibrium distance}
\label{subsec.method}

To determine the equilibrium distance 
of the elastic capsule from the channel center,
we 
considered
two different methods as described in \prettyref{sec:equi_pos}. 
\begin{figure}[h!]
	\includegraphics[width=\linewidth]{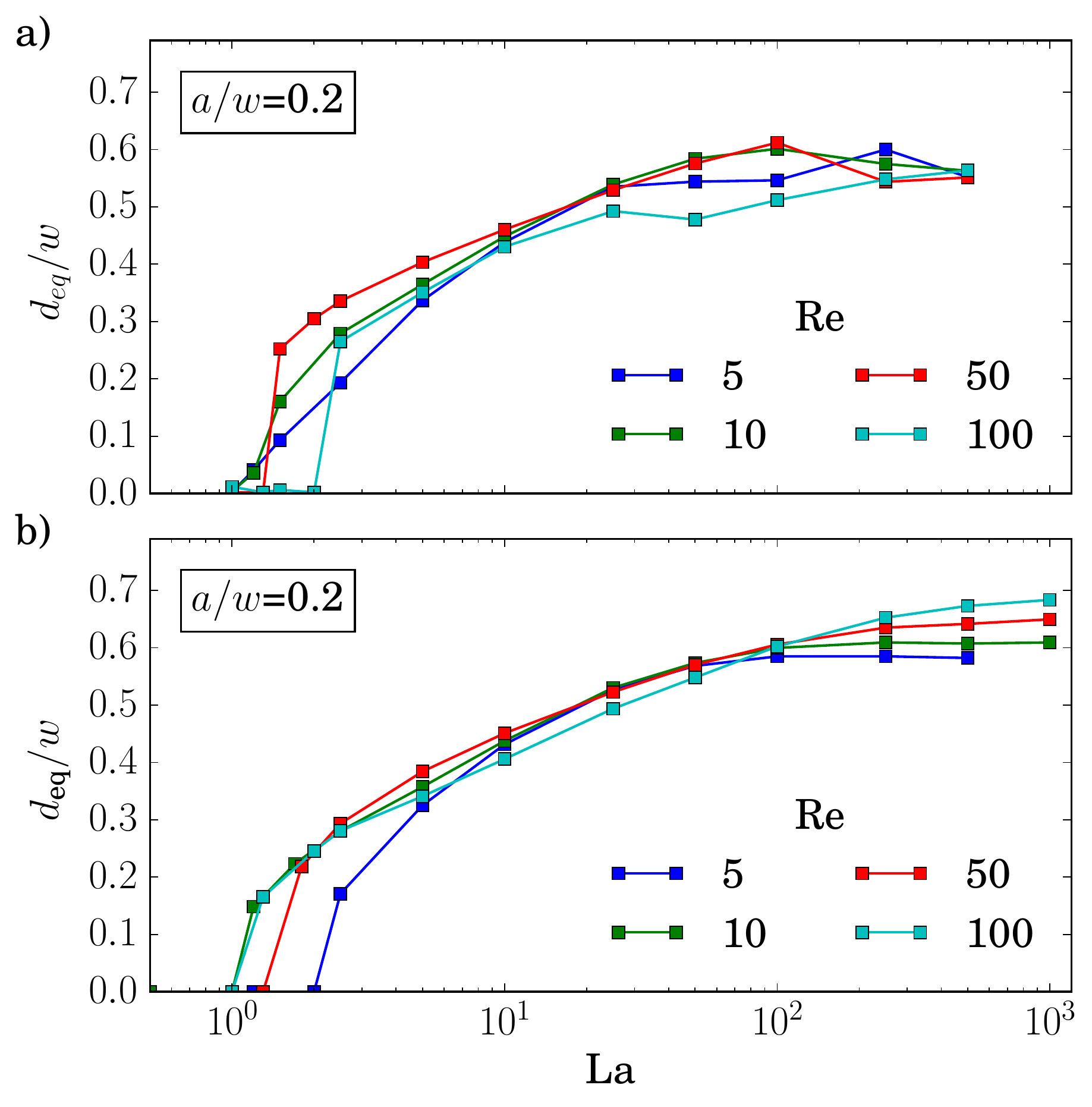}
	\caption{Equilibrium 
	distance from the center
	as a function of the Laplace number 
	determined by
	two different methods. 
	a) Particles are placed in the flow and 
	migrate freely
	to their equilibrium position;
  b)the equilibrium distance is extracted from
    the 
    stable
    fix points 
   of
         the lift-force profiles. Both 
         plots
         agree quite well.
         }
	\label{fig:compare_equi_pos}
\end{figure}

\newpage

\subsection{Green strain tensor}
\label{subsec.cauchy}

\citet{skalak_strain_1973} introduced a model for the strain energy of 
the membrane of a
red blood cell.
It is formulated in 
in the framework
of finite strain theory, which investigates the deformation of a body relative to a 
reference configuration. We denote the reference coordinates by $\vec x$ and the 
coordinates
of the deformed state
by $\vec y$. In the coordinate system,
where the local Jacobi matrix $\partial y_i / \partial x_j$ is diagonal,
the extension ratios $\lambda_1$ and $\lambda_2$ are given by
\begin{equation}
\lambda_1=\frac{\partial y_1}{\partial x_1} \quad
\text{and}
\quad \lambda_2=\frac{\partial y_2}{\partial x_2}.
\end{equation}
Now, deformations
can be described by the Green strain tensor, 
the
components 
of which
are defined as
\begin{equation}
e_{11}=\frac{1}{2}\left(\lambda_1^2-1\right) \quad
\text{and} 
\quad e_{22}=\frac{1}{2}\left(\lambda_2^2-1\right) \, .
\end{equation}
Furthermore, the authors 
introduced the
strain energy function $E_s$ 
of Eq.
\prettyref{eq:skalak} to calculate the Piola-Kirchoff stress tensor
\begin{equation}
S_{ij}=\frac{\partial E_s}{\partial e_{ij}}
\end{equation}
where $S_{ij}$ is the tension in the membrane
in terms of the reference coordinates $\vec x$.
The Piola-Kirchoff stress tensor is related to the Cauchy stress tensor by
\begin{equation}
\sigma=\frac{1}{\lambda_1\lambda_2}S_{kl}\frac{\partial y_i}{\partial x_k}\frac{\partial y_j}{\partial x_l}
\end{equation}
The strain energy is a scalar function and invariant under all local rotations of the two-dimensional membrane about the
normal. Thus, it can only depend on the invariants of the Green strain tensor, which are written as

\begin{eqnarray}
I_1 &= & 2(e_{11}+ e_{22}) = \lambda_1^2+\lambda_2^2-2 \\
I_2 & = & 4 e_{11} e_{22} + I_1 =\lambda_1^2\lambda_2^2-1 \, .
\end{eqnarray}

\subsection{Equilibrium distance for external control force}
\label{sec.external}

\begin{figure}%
	\centering
	\includegraphics[width=.85\linewidth]{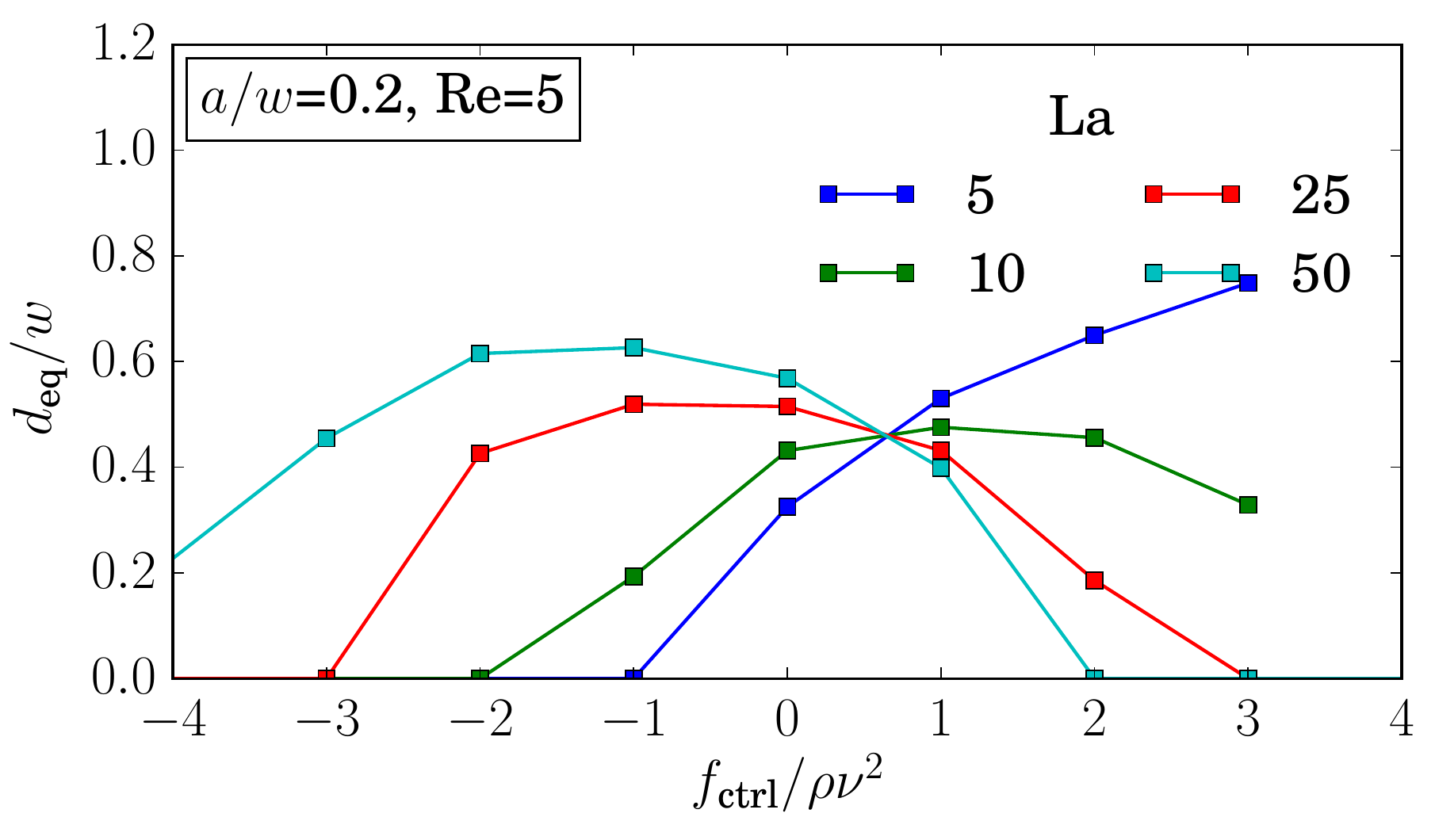}
	\includegraphics[width=.85\linewidth]{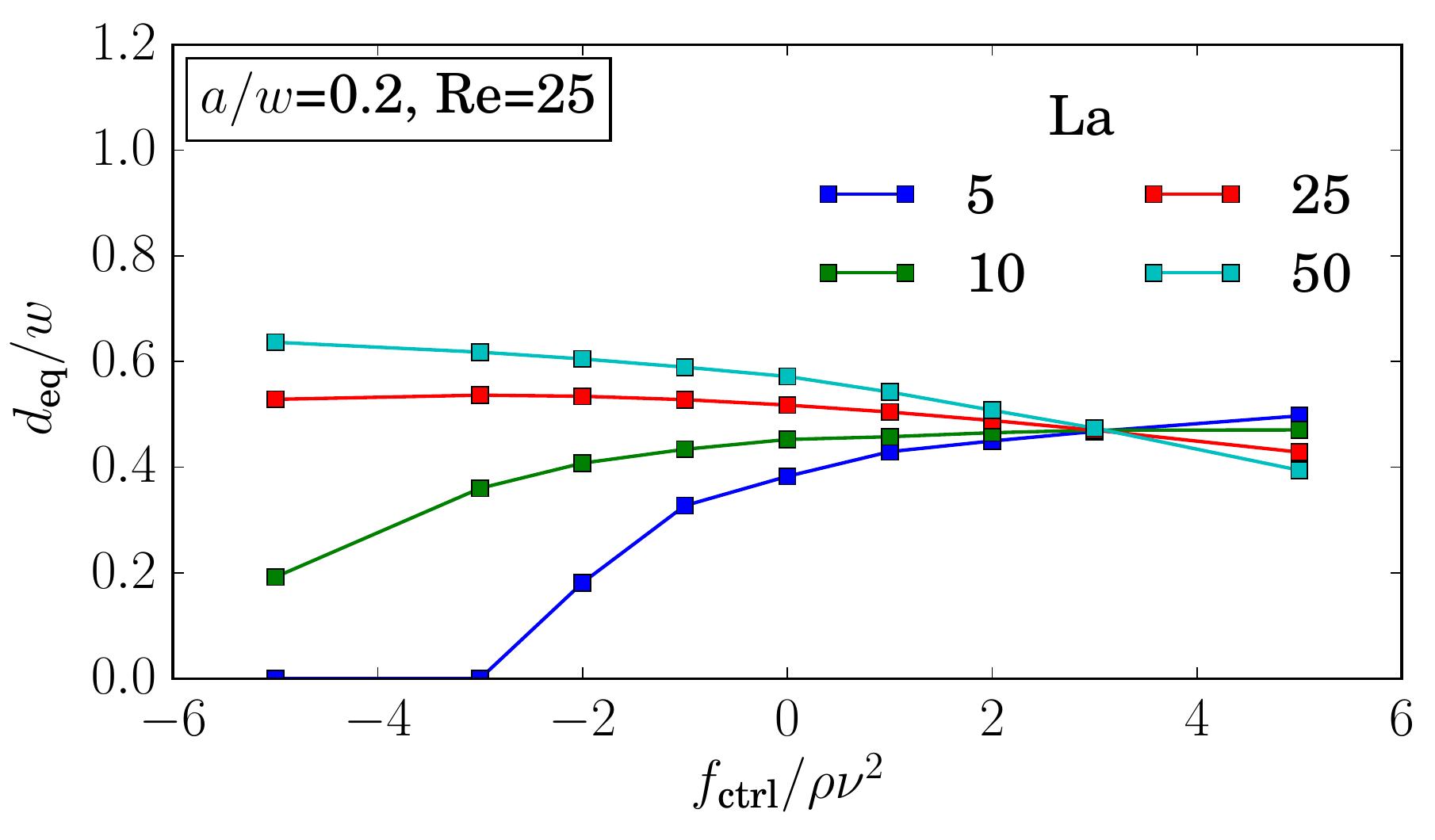}
	\includegraphics[width=.85\linewidth]{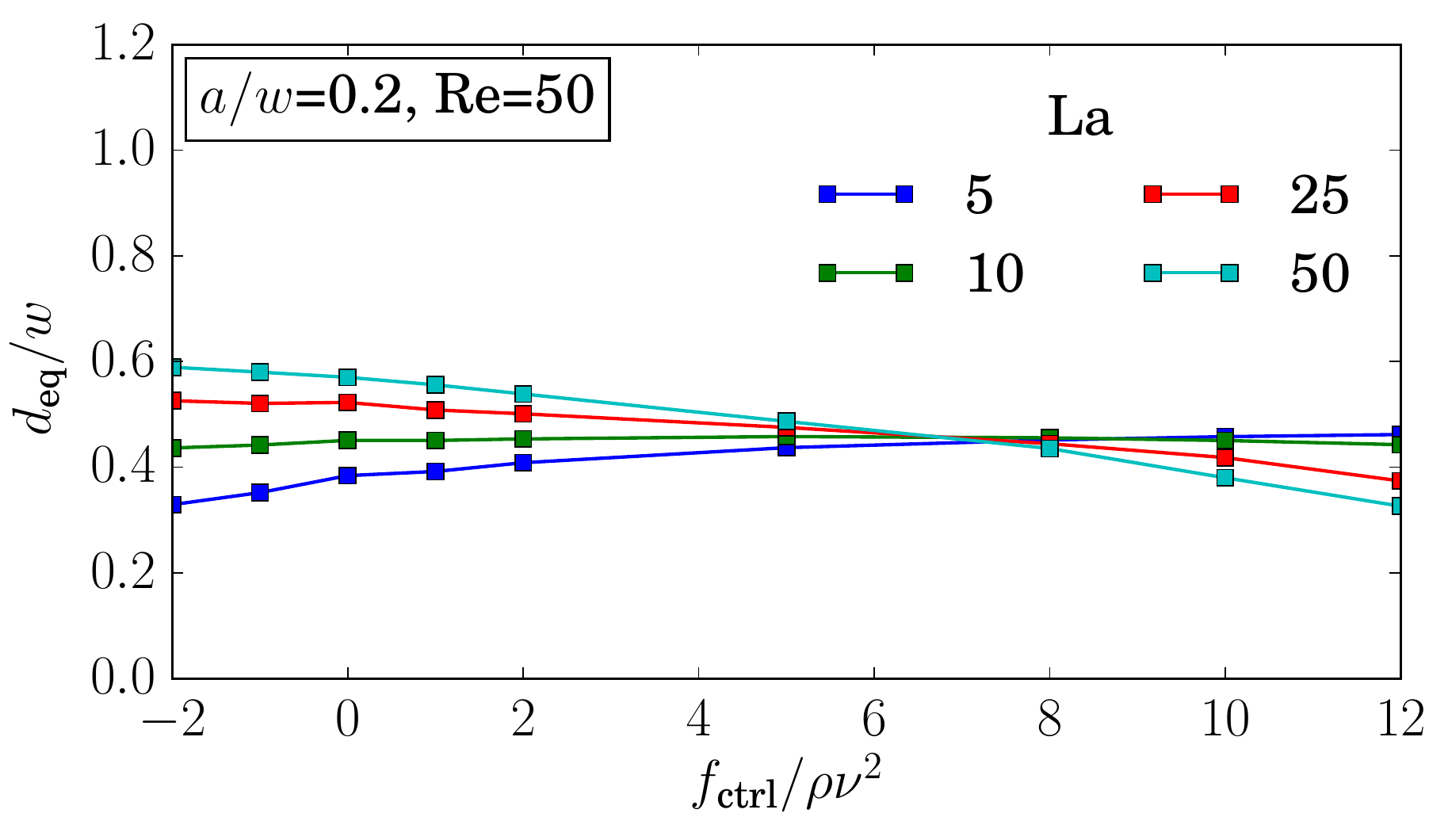}
	\includegraphics[width=.85\linewidth]{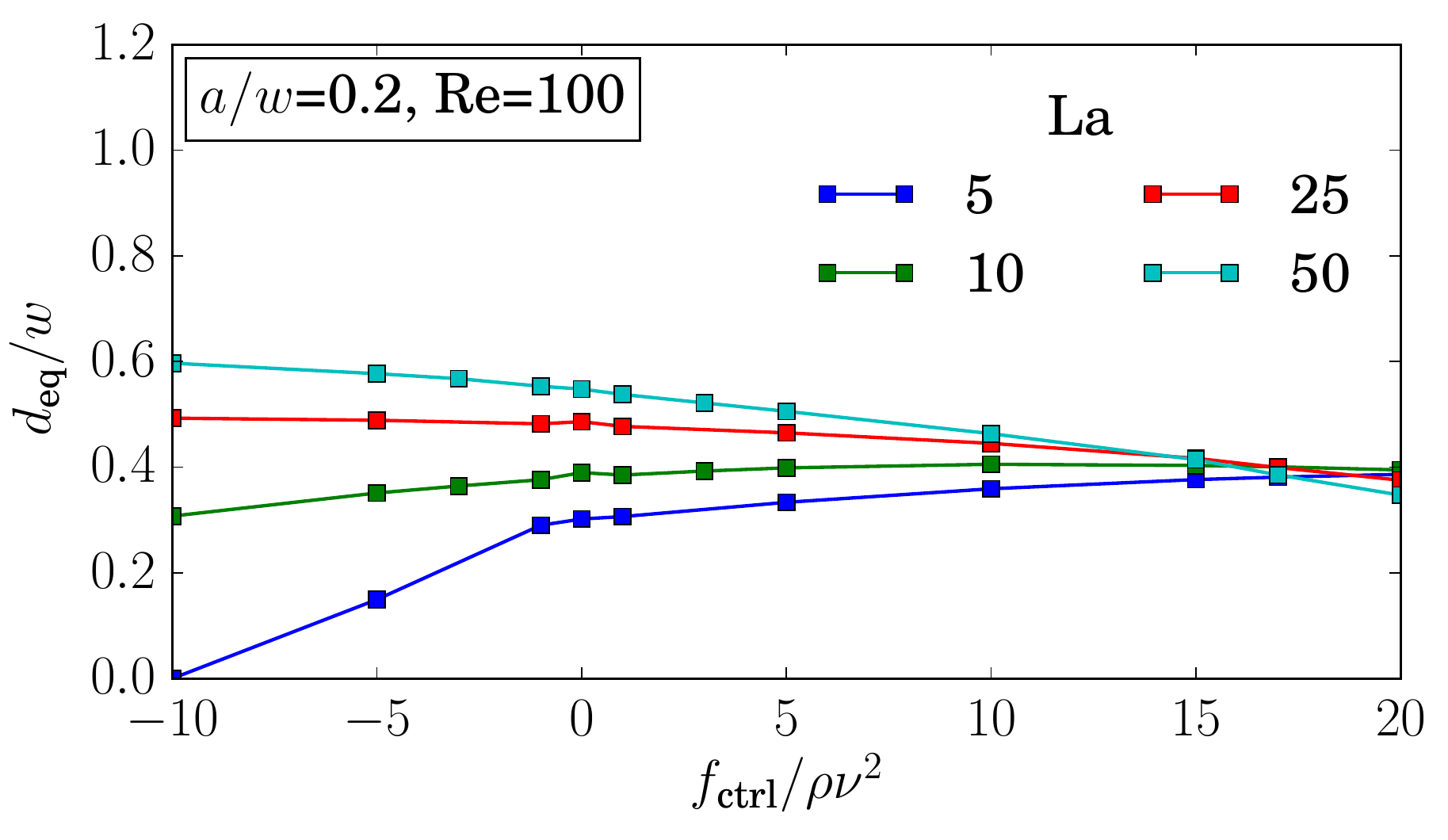}
	\caption{The equilibrium distance $d_{\mathrm{eq}}$
         plotted versus control force $f_{\mathrm{ctrl}}$ for different Laplace numbers $\la$ and Reynolds numbers $\re = 5$, 
         25, 50, 100 for particle radius $a=0.2 w$.
}
	\label{fig:control_reynolds}
\end{figure}

Figure\ \ref{fig:control_reynolds} shows the equilibrium distance plotted versus the control force for different
Laplace numbers and Reynolds numbers. The curves for different $\la$ intersect in one point. This is valid for all
Reynolds numbers. With increasing $\re$ the intersection point moves towards higher forces while the equilibrium
distance only varies little.

\begin{figure}%
	\centering
	\includegraphics[width=.85\linewidth]{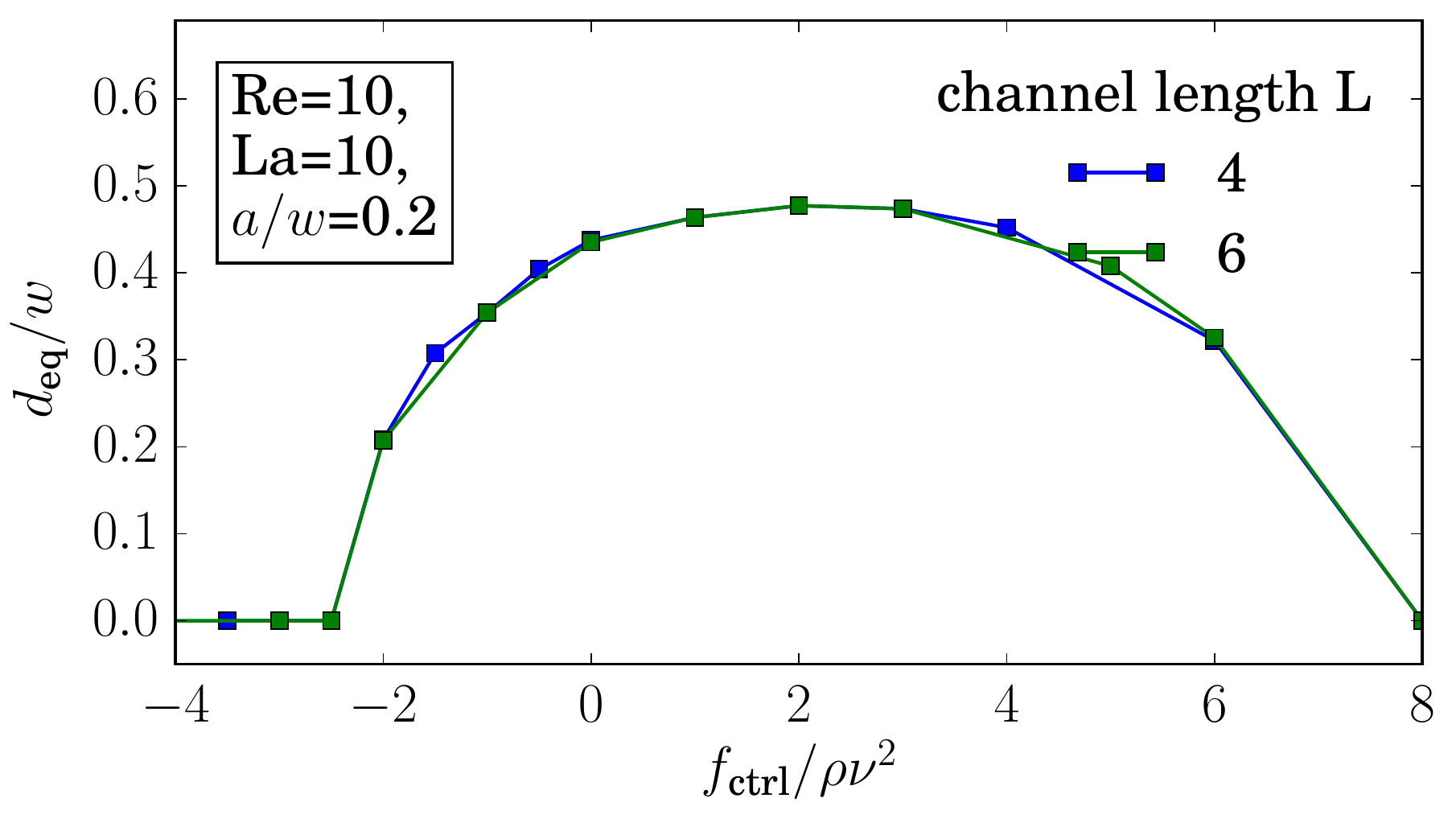}
	\includegraphics[width=.85\linewidth]{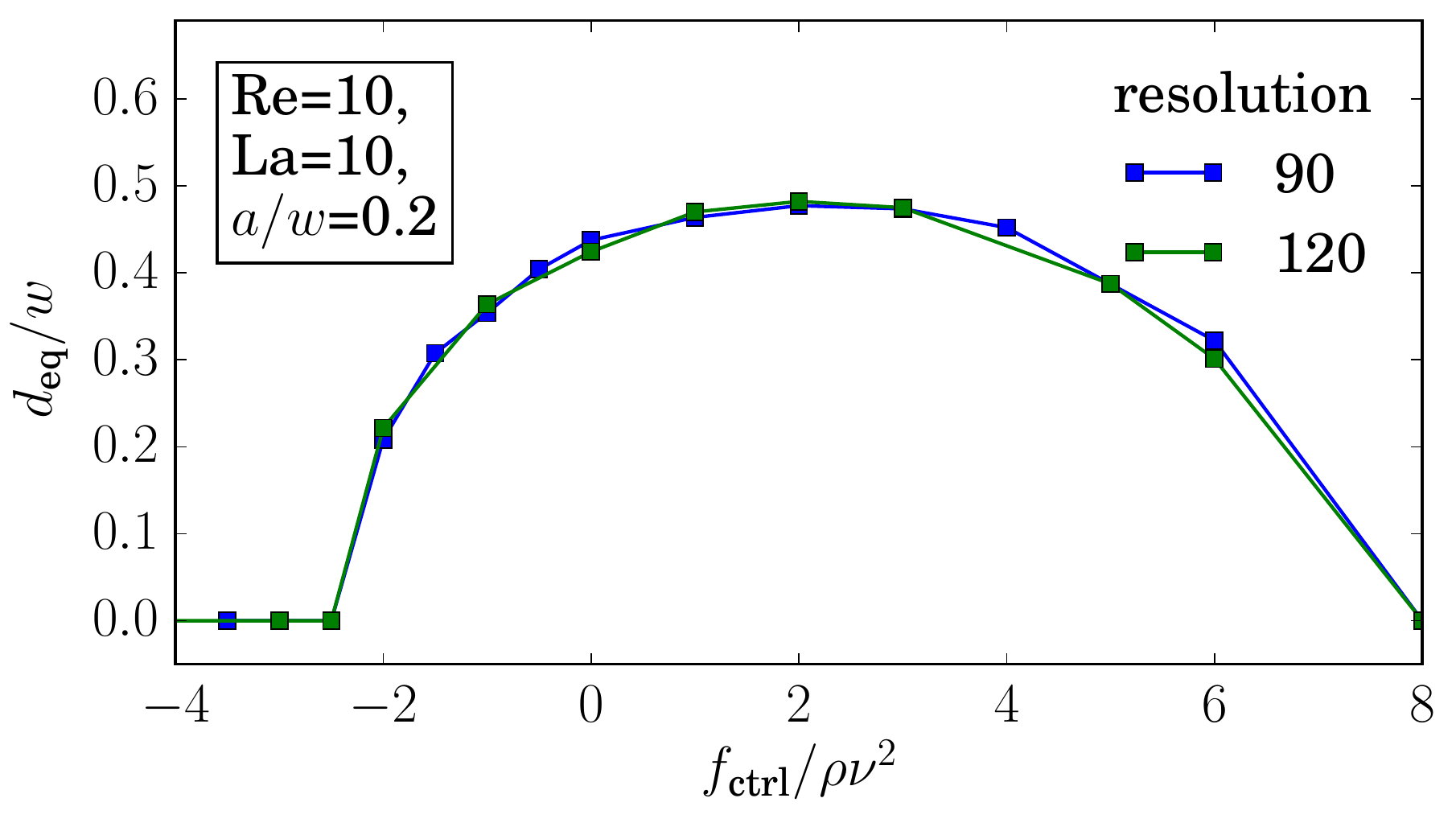}
	\includegraphics[width=.85\linewidth]{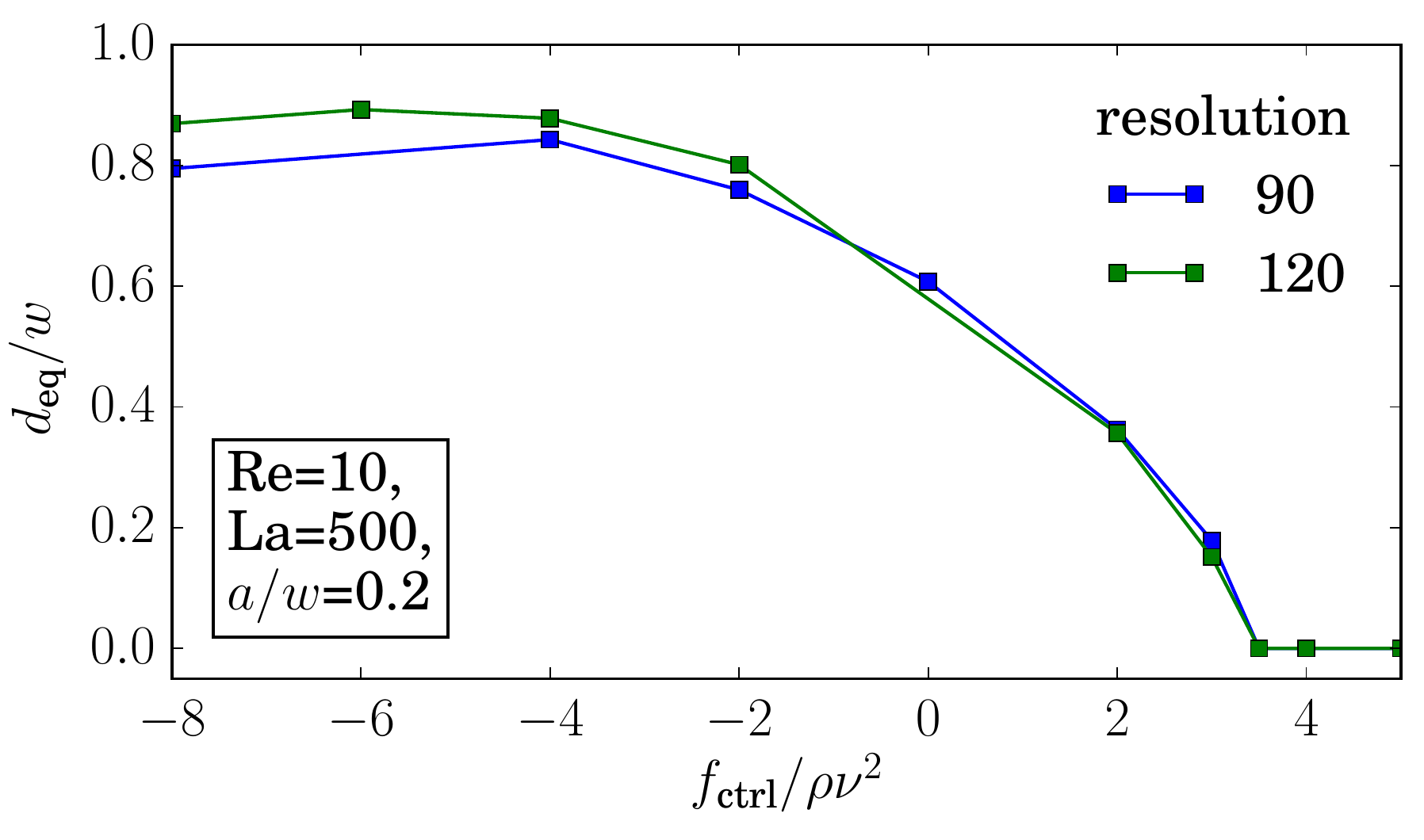}
	\caption{The equilibrium distance $d_{\mathrm{eq}}$ plotted versus control force 
	for different channel lengths (top) and different resolutions of the lattice-Boltzmann grid for $\la=10$ (middle)
	and $\la=500$ (bottom).
}
	\label{fig:check_control}
\end{figure}

As demonstrated in Fig.\ \ref{fig:check_control}, the graphs for the equilibrium distance plotted versus control
force is independent of the channel length $L$, the resolution of the lattice-Boltzmann grid. Only for rigid particles
($\la=500$), one observes small deviations, when the particles are located close to the walls.

\end{document}